\documentclass[useAMS,usenatbib]{mn2e}
\usepackage{amsmath}
\usepackage{times}
\usepackage{ctable,longtable}
\usepackage{graphicx}
\usepackage[colorlinks=true, linkcolor=blue, citecolor=blue, urlcolor=blue]{hyperref}
\title[The dust energy balance in the edge-on spiral galaxy NGC\,4565]{The dust energy balance in the edge-on spiral galaxy NGC\,4565}

\author[I. De Looze et al.]{
Ilse De Looze$^1$, Maarten Baes$^1$, George~J. Bendo$^2$, Laure Ciesla$^3$, Luca Cortese$^4$, 
\newauthor 
Gert De Geyter$^1$, Brent Groves$^5$, M{\'e}d{\'e}ric Boquien$^3$, Alessandro Boselli$^3$, Lena Brondeel$^{1,6}$, 
\newauthor 
Asantha Cooray$^7$, Steve Eales$^8$, Jacopo Fritz$^1$, Fr{\'e}d{\'e}ric Galliano$^{9}$, Gianfranco Gentile$^1$, 
\newauthor 
Karl~G. Gordon$^{10,1}$, Sacha Hony$^{9}$, Ka~H. Law$^{10}$, Suzanne~C. Madden$^{9}$, Marc Sauvage$^{9}$, 
\newauthor
Matthew~W.~L. Smith$^8$, Luigi Spinoglio$^{11}$ and Joris Verstappen$^1$ \\
$^1$ Sterrenkundig Observatorium, Universiteit Gent, Krijgslaan 281 S9, B-9000 Gent, Belgium \\
$^2$ UK ALMA Regional Centre Node, Jodrell Bank Centre for Astrophysics, School of Physics and Astronomy, \\
University of Manchester, Oxford Road, Manchester M13 9PL, United Kingdom \\
$^3$ Laboratoire dÕAstrophysique de Marseille, UMR6110 CNRS, 38 rue F. Joliot-Curie, F-13388 Marseille, France \\
$^4$ European Southern Observatory, Karl Schwarzschild Str. 2, 85748 Garching bei M{\"u}nchen, Germany \\
$^5$ Max Planck Institute for Astronomy, K{\"o}nigstuhl 17, D-69117 Heidelberg, Germany \\
$^6$ Newtec, Laarstraat 5, B-9100 Sint-Niklaas, Belgium \\
$^7$ Department of Physics \& Astronomy, University of California, Irvine, CA 92697, USA \\
$^8$ School of Physics and Astronomy, Cardiff University, The Parade, Cardiff, CF24 3AA, UK \\
$^{9}$ Laboratoire AIM, CEA/DSM - CNRS - Universit{\'e} Paris Diderot, Irfu/Service d'Astrophysique, CEA Saclay, 91191 Gif-sur-Yvette, France \\
$^{10}$ Space Telescope Science Institute, 3700 San Martin Drive, Baltimore, MD 21218 \\
$^{11}$ Istituto de Astrofisica e Planetologia Spaziali (IAPS) - Istituto Nazionale di Astrofisica (INAF), Via del Fosso del Cavaliere 100, I-00133 Roma, Italy} 

\begin{document}

\date{Received}

\pagerange{\pageref{firstpage}--\pageref{lastpage}} \pubyear{2012}

\maketitle

\label{firstpage}

\begin{abstract}
We combine new dust continuum observations of the edge-on spiral galaxy NGC\,4565 in all \textit{Herschel}/SPIRE (250, 350, 500\,$\mu$m) wavebands, obtained as part of the \textit{Herschel} Reference Survey, and a large set of ancillary data (Spitzer, SDSS, GALEX) to analyze its dust energy balance.
We fit a radiative transfer model for the stars and dust to the optical maps with the fitting algorithm FitSKIRT.
To account for the observed UV and mid-infrared emission, this initial model was supplemented with both obscured and unobscured star-forming regions.
Even though these star-forming complexes provide an additional heating source for the dust, the far-infrared/submillimeter emission long wards of 100\,$\mu$m is underestimated by a factor of 3-4. This inconsistency in the dust energy budget of NGC\,4565 suggests that a sizable fraction (two-thirds) of the total dust reservoir ($M_{d}$ $\sim$ 2.9 $\times$ 10$^{8}$ $M_{\odot}$) consists of a clumpy distribution with no associated young stellar sources. The distribution of those dense dust clouds would be in such a way that they remain unresolved in current far-infrared/submillimeter observations and hardly contribute to the attenuation at optical wavelengths. 
More than two-thirds of the dust heating in NGC\,4565 is powered by the old stellar population, with localized embedded sources supplying the remaining dust heating in NGC\,4565.
The results from this detailed dust energy balance study in NGC\,4565 is consistent with that of similar analyses of other edge-on spirals. 
\end{abstract}


\section{Introduction}
Panchromatic radiative transfer 2D and 3D modelling of galaxies provide a powerful tool to analyze the characteristics of dust in galaxies (i.e. optical properties, distribution, clumpiness, etc.) in a self-consistent way (e.g. \citealt{Xilouris1999,Popescu,2011A&A...527A.109P,Alton,2008A&A...490..461B,2010A&A...518L..39B,2011ApJ...741....6M,2012MNRAS.419..895D,2012arXiv1204.2936H}). From optical observations, the properties and spatial distribution of stars and dust can be constrained using a radiative transfer code to model the propagation of stellar light and its interaction with dust particles in a galaxy.
In a second step, the dust emission predicted from the radiative transfer simulations is compared to the observed thermal dust re-emission at infrared/submillimeter wavelengths.
Such a complementary study imposes that dust features can easily be identified from optical as well as infrared observations. This requirement has limited the number of galaxies for which detailed dust energy balance studies have been attempted in the past. 
Edge-on spirals are considered ideal cases for those complementary studies since projection effects allow to resolve the dust distribution vertically (e.g. \citealt{2001A&A...372..775M,Alton,Dasyra,2008A&A...490..461B,2010A&A...518L..39B,2011A&A...527A.109P,2012MNRAS.419..895D}). 
Unfortunately, the large inclination angle impedes the characterization of substructures in the dust organization and the identification of heavily obscured star-forming regions. Only the application of complex 3D models in conjunction with competitive fitting algorithms have been shown to successfully characterize clumpiness and asymmetries in the stellar and dust distribution of highly inclined objects (e.g. \citealt{2012ApJ...746...70S}).

Dust energy balance studies of individual edge-on galaxies reveal an inconsistency between the predicted FIR/submm fluxes of radiative transfer models and the observed emission in those wavebands (e.g. \citealt{Popescu,2001A&A...372..775M, Alton, Dasyra, 2010A&A...518L..39B, 2011A&A...527A.109P,2012MNRAS.419..895D,2012arXiv1204.2936H}). Although radiative transfer models might successfully explain the observed optical attenuation, the modelled dust emission underestimates the observed thermal dust re-emission by a factor of 3-4.
In order to reconcile the results of the radiative transfer models with the observations, two scenarios have been proposed: either a significant underestimation of the FIR/submm dust emissivity has been argued (e.g. \citealt{Alton, Dasyra,2011ApJ...741....6M}) or, alternatively, the distribution of a sizable fraction of dust in clumps or a second inner dust disk having a negligible attenuation on the bulk of the starlight (e.g. \citealt{Popescu,2001A&A...372..775M,2008A&A...490..461B,2012MNRAS.419..895D,2011ApJ...741....6M}). However, a general consensus on the origin of the energy balance problem has not yet been achieved. 

Until recently, observations of dust in galaxies were hampered by the poor resolution and limited wavelength coverage of infrared instrumentation. 
The first far-infrared space satellites (e.g. IRAS, ISO, \textit{Spitzer}, \textit{Akari}) only covered mid- and far-infrared wavelengths up to 200\,$\mu$m. Most ground-based submillimeter/mm instrumentation either only start operating from 850\,$\mu$m onwards (e.g. LABOCA, IRAM) or suffer severely from the earth's opacity at shorter submm wavelengths (450\,$\mu$m, e.g. SABOCA, SCUBA). This coarse sampling of the Rayleigh-Jeans tail of dust emission in most galaxies induced poor constraints on the temperature, content and properties of dust in those objects.
Recently, this large void in wavelength coverage was bridged thanks to the launch of the \textit{Herschel} Space Observatory \citep{2010A&A...518L...1P}, offering the opportunity to trace the dust emission from galaxies in a broad wavelength domain ranging from 55 to 672\,$\mu$m. The high spatial resolution achieved by \textit{Herschel} is providing a large collection of galaxies suitable for detailed dust energy balance analyses, supplemented with the ongoing development of powerful radiative transfer codes.

At a distance of about 16.9 Mpc, the Needle Galaxy (NGC\,4565) is one of the most nearby edge-on spiral galaxies covering an angular scale of $\sim$ 15$\arcmin$ on the sky. Distance measurements for NGC\,4565 range from 9 to 22 Mpc depending on the applied technique. We adopt a value of $D$ $\sim$ 16.9 Mpc throughout this paper, which is the average of the most recent distance measurements from $I$ band surface brightness fluctuations as reported in \citet{2001ApJ...546..681T} and \citet{2003ApJ...583..712J}. All values from the literature referred to in this paper were converted to the assumed distance of 16.9 Mpc.
NGC\,4565 is a Sb spiral, classified as a Seyfert galaxy \citep{1997ApJS..112..391H} and located in the Coma I cloud. Recently its classification as an active galactic nucleus has been confirmed in the mid-infrared with the detection of the [Ne V] lines with \textit{Spitzer} \citep{2010AJ....140..753L}, which are among the strongest indicators of AGNs (see e.g. \citealt{2010ApJ...709.1257T}). The stellar geometry in NGC\,4565 has been studied extensively at optical and NIR wavelengths \citep{1979A&AS...38...15V,1980PASJ...32..197H,1981A&A....95..105V,1982ApJS...50..421J,1986A&A...167L..21D,1996AJ....112..114R,1997A&A...325..915N,2002AJ....123.1364W,2005AJ....129.1331S,2010MNRAS.401..559M,2010ApJ...715L.176K}. 
From studies of the absorption layer in NGC\,4565, \citet{1995PASJ...47...17O} claim a dust distribution in a ring-like structure and an inner cut-off radius along the major axis at $\sim$ 130$\arcsec$. At radii smaller than the cut-off radius dust obscuration effects seem to be absent, either due to a deficiency of dust or the distribution of a central dust component preventing the identification of dust attenuation from optical data \citep{1995PASJ...47...17O}.
Studies of residual $J$ and $K$ band images obtained after fitting the stellar component and subtracting those models from the near-infrared images in \citet{1996AJ....112..114R} result in estimates for the dust scale length $\sim$ 66$\arcsec$ and scale height $\sim$ 5$\arcsec$ in NGC\,4565, which corresponds to about  60 and 50$\%$ of the stellar disk in the $K$ band. 

In addition to these optical/NIR studies of the dust component, the properties and geometry of the dust in NGC\,4565 have been analyzed as well from infrared/submillimeter observations \citep{1987A&A...181..225W,1992ApJ...394..104E,1996AJ....112..114R,1996A&A...310..725N,Alton,2010ApJ...715L.176K,2010AJ....140..753L}. In the mid-infrared IRAC 8$\mu$m waveband, \citet{2010ApJ...715L.176K} and \citet{2010AJ....140..753L} identify a ring of prominent polycyclic aromatic hydrocarbon emission from dust at the same position as the molecular gas ring at radii of $\sim$ 80-100$\arcsec$ \citep{1994PASJ...46..147S,1996A&A...310..725N}.
The first far-infrared study on dust emission in NGC\,4565 based on IRAS data calculate a total infrared luminosity $L_{\text{FIR}}$ $\sim$ 3 $\times$ 10$^{9}$ L$_{\odot}$ \citep{1987A&A...181..225W}, which compares well to the dust emission from our Galaxy. The dust emission in NGC\,4565 however is more extended compared to the dust distribution in our Galaxy \citep{1987A&A...181..225W}. Based on estimates of the infrared excess, \citet{1987A&A...181..225W} argue that only one quarter of the dust reservoir in NGC\,4565 is heated by embedded stars in molecular clouds with the remainder of the dust heating power provided by the diffuse interstellar radiation field.

Observations at FIR wavelengths (100, 160, 200\,$\mu$m) with the Kuiper Airborne Observatory \citep{1992ApJ...394..104E} find a two-component dust model with a cold dust  reservoir ($T_{d}$ $\sim$ 20 K) distributed in an exponential disk and a warmer dust component ($T_{d}$ $\sim$ 30 K) spatially coinciding with a bisymmetric spiral pattern in NGC\,4565. 
In spatial correlation with the more extended ring of H{\sc{i}} gas, \citet{1996A&A...310..725N} also found a colder dust component ($\sim$ 15 K) in the outskirts of the galaxy. A second dust reservoir ($T_{d}$ $\sim$ 18 K) was found in the center of NGC\,4565 and in the molecular gas ring at radii of $\sim$ 80-100 $\arcsec$. The central concentration of both dust and molecular gas \citep{1996A&A...310..725N} contradicts the suggested absence of dust in the centre of the galaxy \citep{1995PASJ...47...17O}. 

\citet{Alton} constrain the amount of dust in NGC\,4565 from a complementary study of the attenuation properties of dust in the optical and the thermal dust re-emission at 1.2 mm. 
From optical constraints, they construct a radiative transfer model for the stars and dust in NGC\,4565. Upon comparison of the visual optical depth in their radiative transfer model with thermal continuum radiation in the 1.2 mm waveband, \citet{Alton} infer a dust emissivity at 1.2 mm which is 1.5 times higher than the benchmark, semi-empirical model from \citet{1984ApJ...285...89D} used in the radiative transfer simulation. From their combined study on the dust emissivity in NGC\,4565 and two other nearby spirals, they argue that coagulation of well-ordered dust crystalline grains into amorphous particles in high-density environments can influence the emissivity of dust at submm/mm wavelengths.

The first submillimeter observations of NGC\,4565 were obtained during the first test flight of the 2 m Balloon-borne Large Aperture Submillimeter Telescope (BLAST) in 2005 \citep{2009ApJ...707.1809W}, covering the galaxy in the same wavebands (250, 350, 500\,$\mu$m) as offered by the SPIRE instrument onboard the \textit{Herschel} Space Observatory. Although the resolution in their final images was degraded by the large point-spread functions (PSF) ($\sim$ 186-189$\arcsec$, \citealt{2008ApJ...681..415T}), \citet{2009ApJ...707.1809W} could estimate the scale length of the dust lane $h_{R}$ $\sim$ 150$\arcsec$ from the beam-convolved image at 250\,$\mu$m, which is about twice the value obtained from optical studies \citep{1996AJ....112..114R}.
A modified black-body fit to the BLAST fluxes and additional constraints from ISO (170\,$\mu$m, \citealt{2004A&A...422...39S}) and IRAS (12, 25, 60, 100\,$\mu$m, \citealt{2003AJ....126.1607S}) observations predict a dust reservoir of mass $M_{d}$ $\sim$ 3 $\times$ 10$^{8}$ $M_{\odot}$ (scaled to a distance of 16.9 Mpc) at a temperature $T_{d}$ $\sim$ 16 K for an emissivity index $\beta$ = 2 and a \citet{2007ApJ...657..810D} dust grain composition \citep{2009ApJ...707.1809W}. 

NGC\,4565 was recently observed in all SPIRE wavebands (250, 350 and 500\,$\mu$m) as part of the \textit{Herschel} Reference Survey (HRS, \citealt{2010PASP..122..261B}). In this work, we will analyze the dust characteristics from a spatially resolved study of the \textit{Herschel} infrared/submillimeter photometry and a wealth of ancillary observations.
Section \ref{Obs.sec} reviews the observing and data reduction strategy for the \textit{Herschel} observations and the set of ancillary data used in this study. 
Section \ref{Balance.sec} presents a detailed study of the dust energy balance in NGC\,4565 with a brief introduction to the applied radiative transfer code SKIRT and fitting algorithm FitSKIRT (Section \ref{SKIRT.sec}) and a gradual construction of a self-consistent model that can account for the observed quantities for NGC\,4565 across the multi-wavelength spectrum (Section \ref{RTModel.sec}).
In Section \ref{Discussion.sec}, the results of our radiative transfer modelling procedure are discussed and compared to the dust energy balance, gas-to-dust ratio and dust heating mechanisms in other nearby galaxies.

\section{Observations and data reduction}
\label{Obs.sec}
\subsection{Herschel data (250-500 micron)}
\label{SPIRE.sec}
NGC\,4565 was observed on the 31st of December 2009 (ObsID 0$\times$50002cd8) as part of the \textit{Herschel} Reference Survey (HRS, \citealt{2010PASP..122..261B}), a \textit{Herschel} Guaranteed Time Key Program observing 323 galaxies in the nearby Universe spanning a wide range in morphological type and environment.
Dust continuum maps were obtained in all SPIRE wavebands (centered at 250, 350 and 500\,$\mu$m), covering the galaxy in three cross-linked scans (nominal and orthogonal) at medium scan speed (30$\arcsec$/s).
The final maps cover a square area of 30$\arcmin$ $\times$ 30$\arcmin$, corresponding to an extent of at least 1.5 times the optical isophotal diameter $D_{25}$ in NGC\,4565. 

\begin{table}
\caption{Panchromatic overview of flux densities used in the dust energy balance study of NGC\,4565}
\label{dataptn}
\begin{center}
\begin{tabular}{@{}|l|c|c|c|}
\hline 
Filter & $\lambda$~($\mu$m) & $F_{\nu}$~(Jy) & Ref\footnotemark[1]  \\
\hline \hline 
FUV & 0.15 & 0.008 $\pm$ 0.001  & 1 \\
NUV & 0.23 & 0.015 $\pm$ 0.002 & 1 \\

SDSS u & 0.36 & 0.07 $\pm$ 0.01 & 2 \\
SDSS g & 0.47 & 0.33 $\pm$ 0.03 & 1 \\ 
SDSS r & 0.62 & 0.70 $\pm$ 0.07 & 1 \\
SDSS i & 0.75 & 1.08 $\pm$ 0.10 & 1 \\
SDSS z & 0.89 & 1.45 $\pm$ 0.14 & 2 \\
2MASS J &  1.25 & 2.18 $\pm$ 0.03 & 3 \\
2MASS H &  1.65 & 3.05 $\pm$ 2.51 & 3 \\
2MASS K &  2.20 & 2.51 $\pm$ 0.04 & 3 \\
WISE 1 & 3.4 & 1.54 $\pm$ 0.04 & 4 \\
IRAC 3.6 & 3.6 &  1.39 $\pm$  0.14 & 5 \\
IRAC 4.5 & 4.5 &  0.91 $\pm$ 0.09 & 5 \\
WISE 2 & 4.6 & 0.90 $\pm$ 0.03 & 4 \\
IRAC 5.8 & 5.8 &  1.17 $\pm$ 0.14 & 5 \\
IRAC 8.0 & 8.0 &  1.85  $\pm$ 0.19  & 5 \\
WISE 3 & 12.1 & 2.20 $\pm$ 0.05 & 4 \\
IRAS 12 & 12 &  1.53 $\pm$ 0.23  & 6\\
WISE 4 & 22.2 & 1.60 $\pm$ 0.16 & 4 \\
MIPS 24 & 24 &  1.65 $\pm$ 0.07  & 6 \\
IRAS 25 & 25 &  1.70 $\pm$ 0.26  & 6\\
IRAS 60 & 60 &  9.83 $\pm$ 1.47 &  6\\
MIPS 70 & 70 &  19.3 $\pm$ 1.9 & 6 \\
IRAS 100 & 100 & 47.23  $\pm$ 7.08  & 6 \\
KUIPER 100  & 100 & 52.00 $\pm$ 10.40  & 8 \\ 
MIPS 160 & 160 & 86 $\pm$ 10  & 7 \\
KUIPER 160  & 160 & 62.00 $\pm$ 12.40 & 8 \\ 
KUIPER 200  & 200 & 55.00 $\pm$ 11.00 & 8 \\ 
SPIRE 250 & 250 & 63.24 $\pm$ 1.18 & 9\\
SPIRE 350 & 350 & 31.35 $\pm$ 0.69 & 9\\
SPIRE 500 & 500 & 13.14 $\pm$ 0.29 & 9\\
IRAM 1200 & 1200 & 1.00 $\pm$ 0.10 & 10\\
\hline 
\end{tabular}
\end{center}
\footnotemark[1]{References: (1) \citealt{Cortese}; (2) Cortese (priv. comm.); (3) \citealt{2003AJ....125..525J}; (4) this paper, see Section \ref{OtherIR}; (5) Ciesla et al.\,(in prep.);  (6) \citealt{1988ApJS...68...91R}; (7) \citealt{2012arXiv1202.4629B}; (8) \citealt{1992ApJ...394..104E}; (9) \citealt{Ciesla}; (10) \citealt{1996A&A...310..725N} }\\
\end{table}

The SPIRE data were processed up to Level-1 in the \textit{Herschel} Interactive Processing Environment
(HIPE, version 4.0.1367, \citealt{2010ASPC..434..139O}) with the standard script adapted from the official pipeline (POF5 pipeline.py, dated 8 Jun 2010) as provided by the SPIRE Instrument Control Centre (ICC). 
Instead of the ICC default settings for deglitching, we used the waveletDeglitcher, which was adjusted to mask the sample following a glitch. 
The waveletDeglitcher algorithm was applied a second time after the flux calibration, since this was shown to significantly improve the removal of remaining glitches for the version of the code at that time (currently it is only run once).
Instead of the pipeline default temperature drift correction and median baseline subtraction, we applied a custom method (BriGAdE; Smith et al. in prep.) to remove the temperature drift and bring all bolometers to the same level.
Final SPIRE maps (see Figure \ref{Images_all.pdf}, last three bottom panels) were created with the naive mapper in HIPE with pixel sizes of 6$\arcsec$, 8$\arcsec$, and 12$\arcsec$ at 250, 350, and 500\,$\mu$m, respectively. 
The FWHM of the SPIRE beams have sizes of 18.2$\arcsec$, 24.5$\arcsec$, and 36.0$\arcsec$\footnote[1]{Retrieved from document available at  http://herschel.esac.esa.int/\\ twiki/pub/Public/SpireCalibrationWeb/beam\_release\_note\_v1-1.pdf} at 250, 350, and 500\,$\mu$m (for pixel sizes of 6, 8 and 12$\arcsec$), respectively. 
To update our flux densities to the latest v8 calibration product, the 350\,$\mu$m images are multiplied by a factor of 1.0067 \citep{SPIRE}.
Since the calibration in the standard data reduction pipeline was optimized for point sources, correction factors (0.9828, 0.9834 and 0.9710 at 250, 350 and 500\,$\mu$m) are applied to convert the K4 colour correction factors from point source to extended source calibration \citep{SPIRE}. Appropriate colour correction factors were determined from a single component modified blackbody fit to the Rayleigh-Jeans tail of dust emission (constrained by fluxes from MIPS 160\,$\mu$m and all SPIRE wavebands) for values $\beta$~=~1.0, 1.5 and 2.0 (i.e. the range of spectral indices derived for the HRS sample galaxies, \citealt{2012arXiv1201.2305B}). With the best fit obtained for values $\beta$~=~2 and $T_{\text{d}}$~=~17 K, we apply colour correction factors (1.019,1.009,1.021) to the 250, 350 and 500\,$\mu$m images \citep{SPIRE}, respectively, to correct for the shape of the response function in every filter.
An uncertainty in the calibration of 7$\%$ was assumed \citep{2010A&A...518L...4S,SPIRE}.

From aperture photometry within an elliptical aperture (20$\arcmin$ $\times$ 2.7$\arcmin$) flux densities of 62.06, 31.07 and 12.87 Jy for NGC\,4565 were obtained at 250, 350 and 500\,$\mu$m, respectively \citep{Ciesla}.
Since those fluxes were corrected for extended source calibration, we only need to correct those values with appropriate colour corrections factors (1.019, 1.009, 1.021). This results in final flux densities of 63.24, 31.35 and 13.14 Jy  at 250, 350 and 500\,$\mu$m, respectively. The large ancillary dataset of far-infrared/submillimeter observations allows us to compare those SPIRE fluxes to flux measurements from other space or airborne missions in overlapping wavebands.
\citet{2009ApJ...707.1809W} report fluxes obtained with BLAST at 250\,$\mu$m (37.2 $\pm$ 4.5 Jy), 350\,$\mu$m (21.0 $\pm$ 2.1 Jy ) and 500\,$\mu$m (9.8 $\pm$ 0.9 Jy), which are 40, 32 and 24 $\%$ lower with respect to SPIRE flux measurements (see Table \ref{dataptn}). BLAST observations of NGC\,4565 covered an area of $\sim$ 0.4 deg$^{2}$ centered on the galaxy, sufficient to detect the dust emission from NGC\,4565 in those wavebands. Considering that BLAST fluxes for NGC\,4565 were measured from data taken during the first BLAST test flight in 2005, the images were degraded by the large point-spread function of BLAST05 (FWHM of $\sim$ 186-189$\arcsec$, \citealt{2008ApJ...681..415T}). Prior to aperture photometry, the images were deconvolved following the method outlined in \citet{2008ApJ...681..428C}.
This beam deconvolution might be the cause of the lower BLAST fluxes compared to SPIRE measurements. Alternatively, calibration issues might be responsible for the small off-set between BLAST and SPIRE.
Also Planck measurements \citep{2011A&A...536A...7P} at 350 (24.35 $\pm$ 1.43 Jy) and 550\,$\mu$m (8.12 $\pm$ 0.48 Jy) are about 20$\%$ below our SPIRE measurements. The lower Planck values can easily be explained from the radius applied for aperture photometry on point sources in their first release catalog \citep{2011A&A...536A...7P}. Indeed, an aperture with radius fixed to the FWHM of Planck at 350\,$\mu$m ($\sim$ 4.23$\arcmin$) does not cover the entire galaxy's emission in that waveband (see the upper right panel of Figure 10 in \citealt{Ciesla}). Performing aperture photometry within the Planck radius at 350\,$\mu$m on the SPIRE images results in a flux density $S_{\nu}$ $\sim$ 25.5 $\pm$ 0.9 Jy \citep{Ciesla}, which is in close agreement with the reported Planck value \citep{2011A&A...536A...7P}. 

Figure \ref{Images_all.pdf} (last three bottom panels) shows the \textit{Herschel} maps for NGC\,4565 in the SPIRE 250, 350 and 500\,$\mu$m wavebands. 
Weak warping signatures at the edges of the dust disk can be identified in all \textit{Herschel} dust maps, with the warp on the northwestern side of the galaxy's disk being more pronounced.
These warp features were earlier identified in NGC\,4565 from optical data (e.g. \citealt{1979A&AS...38...15V, 1982ApJS...50..421J,1997A&A...325..915N}), H{\sc{i}} \citep{1976A&A....53..159S,1991Ap&SS.177..465R} and dust continuum \citep{1996A&A...310..725N} observations and are most likely the relicts of a tidal interaction with its neighboring galaxy NGC\,4562 \citep{1982ApJS...50..421J}.  

Aside from the warps on the edges of the galaxy, three peaks in the dust emission can be identified in the SPIRE 250 and 350\,$\mu$m images (see Figure \ref{Images_all.pdf}, second and third bottom panels). The two brightest emission peaks originate from a narrow ring structure centered at a galacto-centric radius of 80-100$\arcsec$, coinciding with a ring of molecular gas in NGC\,4565 \citep{1994PASJ...46..147S,1996A&A...310..725N}. In the center of NGC\,4565 the emission peak corresponds to a central concentration of dust, which was already identified from 1.2 mm observations \citep{1996A&A...310..725N}. Due to the growing size of the SPIRE beam towards longer wavelengths, the three peaks are more difficult to distinguish in the SPIRE 500\,$\mu$m image (see Figure \ref{Images_all.pdf}, fourth bottom panel), only showing a small increase in emission at the edges of the narrow dust ring. 

\setcounter{figure}{0}
\begin{figure*} %
\includegraphics[width=0.98\textwidth]{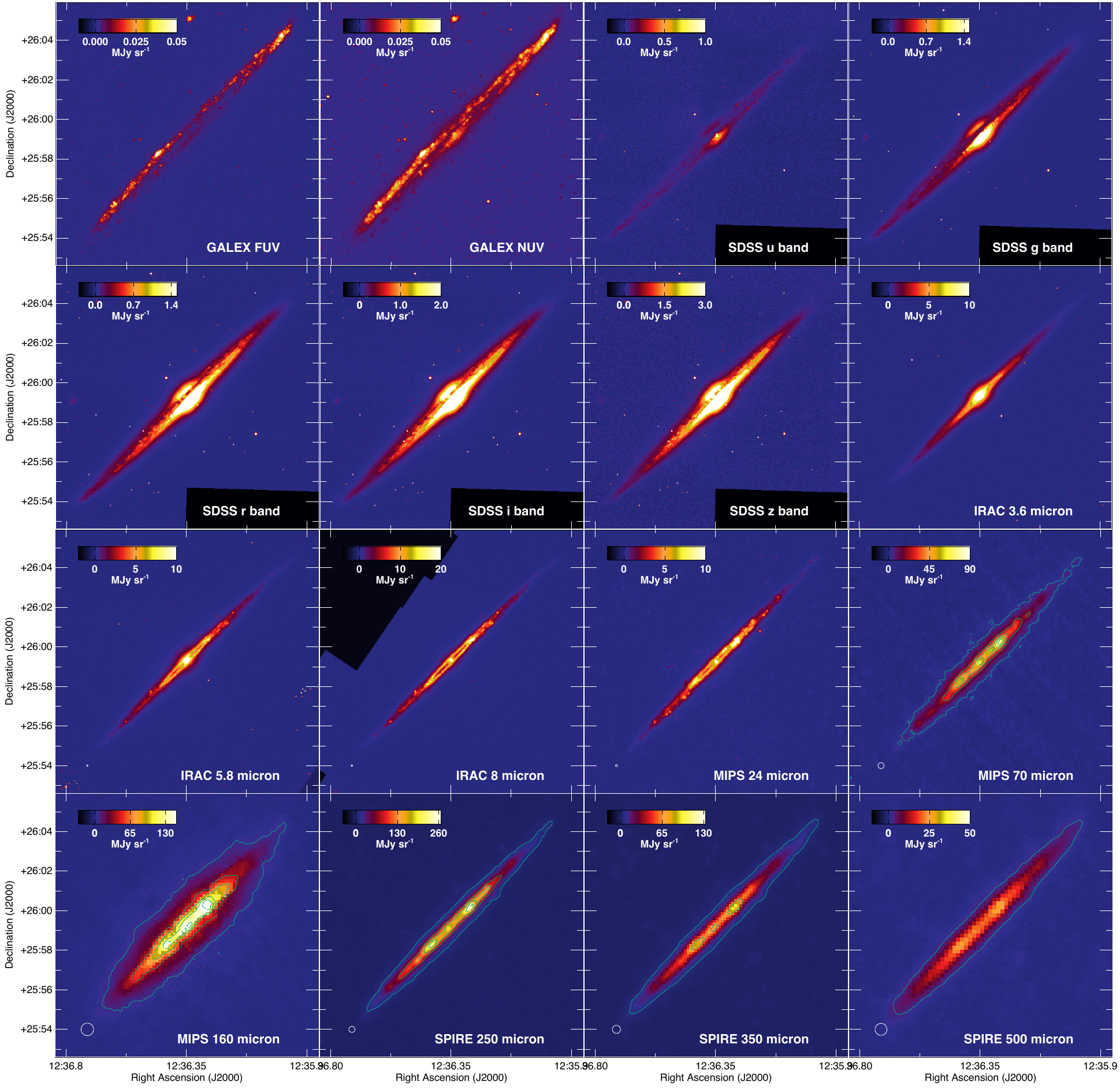} 
\caption{The stellar and dust emission from NGC\,4565 as observed from GALEX $FUV$ and $NUV$, SDSS $ugriz$, \textit{Spitzer} IRAC 3.6, 5.8 and 8.0\,$\mu$m, \textit{Spitzer} MIPS 24, 70 and 160\,$\mu$m and \textit{Herschel} SPIRE 250, 350 and 500\,$\mu$m images. The IRAC 4.5\,$\mu$m map is not explicitly shown here because of the great resemblance with the IRAC 3.6\,$\mu$m image. The beam sizes of all instruments in the infrared/submillimeter wavebands have been indicated as white circles in the bottom left corner of the corresponding panels.}  \label{Images_all.pdf}
\end{figure*}

\setcounter{figure}{1}
\begin{figure*} 
\includegraphics[width=0.98\textwidth]{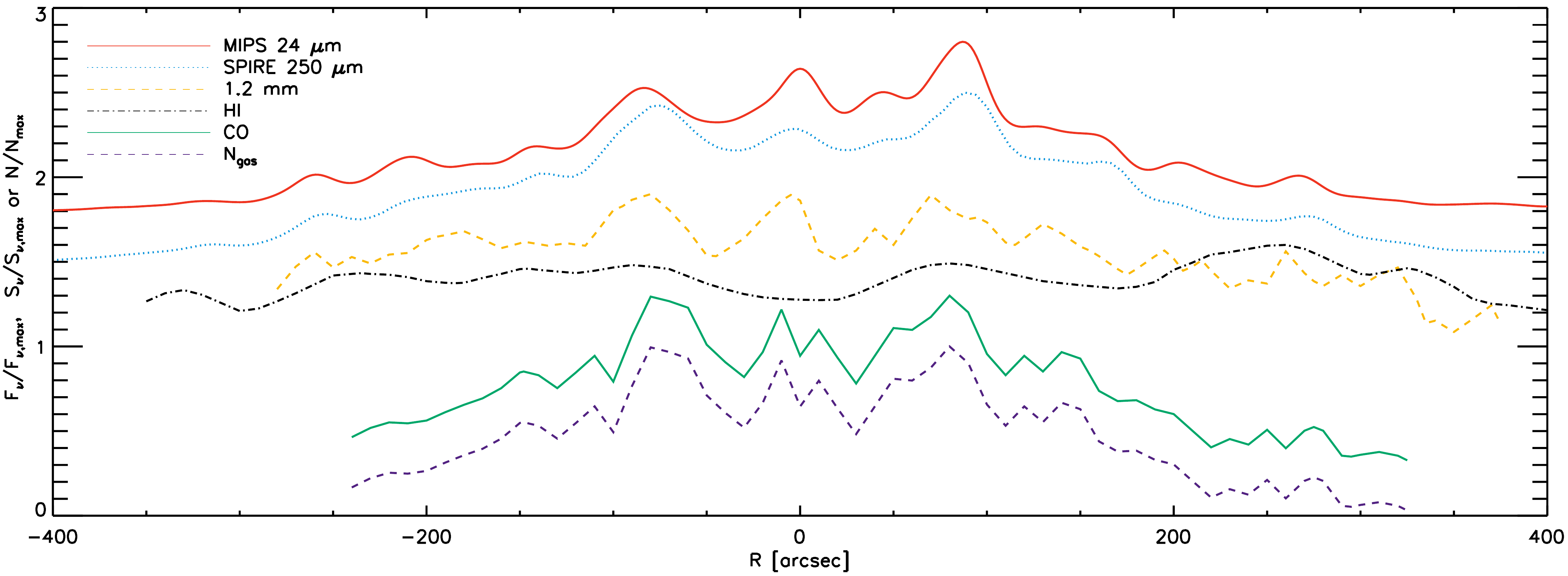} 
\caption{Major-axis profiles of the dust emission in MIPS 24\,$\mu$m (red solid curve), SPIRE 250\,$\mu$m (cyan dotted curve), IRAM 1.2 mm (yellow dashed curve) wavebands, the gas distribution from the H{\sc{i}} (black dashed-dotted curve) and CO (green solid curve) temperature brightnesses and the column density of gas combining the atomic and molecular components in NGC\,4565 (blue dashed curve). The H{\sc{i}} column density could be obtained directly from the H{\sc{i}} temperature brightness ($N_{HI}$~=~1.822 $\times$ 10$^{18}$~cm$^{-2}$~$\int$$T_{b}$($\nu$)d$\nu$). For the calculation of the H$_{2}$ column density from the CO temperature brightness ($N_{H_{2}}$~=~$X_{CO}$~$\int$$T_{b}$($\nu$)d$\nu$), we assumed a factor $X_{CO}$ representative for the H$_{2}$-to-CO fraction in our own Galaxy ($X_{CO}$ $\sim$ 2.3 $\times$ 10$^{20}$ cm$^{-2}$ (K km s$^{-1}$)$^{-1}$, \citealt{1988A&A...207....1S}). All data were smoothed to a resolution of 20$\arcsec$. For ease of reference, all profiles were normalized by their maximum value. Those normalized profiles were then given an offset in the vertical direction to facilitate a mutual comparison.}  \label{Major_gas_dust.pdf}
\end{figure*}

Figure \ref{Major_gas_dust.pdf} compares the distribution of dust in the SPIRE 250\,$\mu$m image with the gas distribution from integrated H{\sc{i}} and CO line intensities \citep{1996A&A...310..725N}, the total gas column density and dust emission profiles along the major-axis from 24\,$\mu$m and 1.2\,mm dust continuum observations, smoothed to a common resolution of 20$\arcsec$. Based on these major axis profiles, we can identify a depression up to the 80$\arcsec$ radius from the center in the H{\sc{i}} profile and a more or less constant column density profile in the outskirts of the galaxy from that radius onwards. The CO profile in NGC\,4565 indicates the presence of a large molecular gas reservoir in the inner regions, distributed in a central disk and narrow ring with peak density around 80$\arcsec$, where also the inner depression in H{\sc{i}} density ends.
The SPIRE 250\,$\mu$m dust emission profile along the major axis correlates well with the molecular gas density profile (see also \citealt{1987A&A...181..225W}, \citealt{1996AJ....112..114R}, \citealt{2010ApJ...715L.176K} and \citealt{2010AJ....140..753L}) as well as the column density of the combined H{\sc{i}} and H$_{2}$ gaseous components. The latter column density profile seems dominated by the peaks in the molecular gas component, rather than the smooth H{\sc{i}} distribution across the plane of the galaxy. In contrast to the CO profile reported in \citet{1996A&A...310..725N}, the CO observations from \citet{1994PASJ...46..147S} identified a more asymmetric CO distribution with the majority of the molecular gas reservoir residing on the northwest side of the galaxy. The dust emission from the colder dust reservoir in the SPIRE 250\,$\mu$m waveband however does not show a tendency for asymmetry. 

\subsection{Spitzer data (3.6-160 micron)}
\label{Spitzer.sec}
NGC\,4565 was observed with the IRAC and MIPS instruments onboard the \textit{Spitzer} Space Observatory as part of the programs \textit{Brown Dwarf Galaxy Haloes} (PI: Giovanni Fazio; AOR key: 3627776) and \textit{The formation of dust lanes in nearby, edge-on disk galaxies} (PI: Roelof S. de Jong; AOR key 14481408), respectively. IRAC (3.6, 4.5, 5.8, 8.0\,$\mu$m) and MIPS (24, 70, 160\,$\mu$m) data were retrieved from the \textit{Spitzer} archive and reduced following the procedure outlined in Ciesla et al.(in prep.) and \citet{2012arXiv1202.4629B}. Although \textit{Spitzer} IRAC data of NGC\,4565 were already used in the analyses of \citet{2010ApJ...715L.176K} and \citet{2010AJ....140..753L}, the maps presented in this work were processed independently.
Final IRAC images have pixel sizes of 0.6$\arcsec$, whereas MIPS images were processed to final maps with pixel sizes of 1.5$\arcsec$, 4.5$\arcsec$ and 9.0$\arcsec$ at 24, 70 and 160\,$\mu$m, respectively. The IRAC beam has an almost uniform size in all wavebands (mean FWHM $\sim$ 1.7$\arcsec$, 1.7$\arcsec$, 1.9$\arcsec$ and 2.0$\arcsec$ at 3.6, 4.5, 5.8 and 8\,$\mu$m, \citealt{2004ApJS..154...10F}). The FWHM of the MIPS beam varies from 6$\arcsec$ at 24\,$\mu$m, 18$\arcsec$ at 70\,$\mu$m to 38$\arcsec$ at 160\,$\mu$m \citep{2007PASP..119..994E,2007PASP..119.1019G,2007PASP..119.1038S}.

To determine flux densities for NGC\,4565 from the processed IRAC images, stars in the surrounding field were masked to obtain realistic estimates for the background emission (Ciesla et al. in prep.). 
Once the images were background subtracted, flux measurements were derived within elliptical aperture matching the shape of the galaxy and encompassing all emission (see Table \ref{dataptn} and Ciesla et al. in prep.). Similarly, the identification and removal of fore- and background emission from MIPS data required some efforts before fluxes could be determined from the background subtracted MIPS images (\citealt{2012arXiv1202.4629B}, see Table \ref{dataptn}).
Uncertainties for IRAC and MIPS photometry measurements include uncertainties owing to the calibration, background noise and  map making. For IRAC, the calibration is assumed to be accurate within 1.8, 1.9, 2.0 and 2.1\,$\%$ in IRAC 3.6, 4.8, 5.8 and 8.0\,$\mu$m wavebands \citep{2005PASP..117..978R}. For MIPS, calibration uncertainties of 4, 10 and 12 $\%$ were reported for the 24, 70 and 160\,$\mu$m wavebands \citep{2007PASP..119..994E,2007PASP..119.1019G,2007PASP..119.1038S}.

MIPS fluxes at 24 and 70\,$\mu$m compare reasonably well with IRAS observations of NGC\,4565 at 25 and 60\,$\mu$m (see Table \ref{dataptn}). The longer wavelength MIPS observations at 160\,$\mu$m are furthermore consistent within 30$\%$ with data from the Kuiper Airborne Observatory (KAO) in the same waveband (see Table \ref{dataptn}).

Figure \ref{Images_all.pdf} (last panel on the second row) shows the stellar emission in the IRAC 3.6\,$\mu$m image. Dust emission from IRAC (5.8 and 8\,$\mu$m) and MIPS observations are presented in Figure \ref{Images_all.pdf} (third row and first panel on fourth row). 
The emission in the IRAC 3.6 and 4.5\,$\mu$m wavebands is dominated by the stellar emission from the bulge in NGC\,4565. The stellar emission from the bulge is also discernible in the IRAC 5.8\,$\mu$m image.
At 5.8 and 8\,$\mu$m, emission from polycyclic aromatic hydrocarbons (PAHs) and very small grains (VSGs) distributed in a ring centered at 80-100$\arcsec$ dominates. 
The presence of a hot dust component heated predominantly by star formation is also clearly detected in the MIPS 24\,$\mu$m image. Besides this ring of hot dust, the central region of NGC\,4565 is also a prominent emission source in the IRAC 5.8, 8\,$\mu$m and MIPS 24\,$\mu$m wavebands. This central dust concentration in NGC\,4565 is most likely heated by the AGN, since the inner 80$\arcsec$ region of NGC\,4565 seems to lack any star formation activity (see Section \ref{GALEXoptical.sec}). 
Owing to the coarser resolution at 70 and 160\,$\mu$m, the dust emission features are not as easily identified compared to shorter wavelength data. To better contrast the structures in the far-infrared dust emission, we overlaid the contours of dust emission on those far-infrared MIPS data. The contours show three peaks in the far-infrared dust emission of NGC\,4565, similar to the peaks identified from the \textit{Herschel} data, with prominent emission from a ring at 80-100$\arcsec$ and a central dust concentration.
Also the dust warps are manifested on the edges of the dust disk, in particular on the northwestern side, in both IRAC (5.8, 8\,$\mu$m) and all MIPS wavebands.

In a similar way as for the SPIRE observations at 250\,$\mu$m, the resolution in the MIPS 24\,$\mu$m map was degraded to 20$\arcsec$, allowing an immediate comparison with the distribution of atomic and molecular gas along the major axis of NGC\,4565 (see Figure \ref{Major_gas_dust.pdf}). With the shape of the PSF in the MIPS 24\,$\mu$m filter deviating from a Gaussian distribution, appropriate kernels were constructed following the procedure outlined in \citet{2012MNRAS.419.1833B} to degrade the resolution in the MIPS 24\,$\mu$m image.
With the 24\,$\mu$m emission tracing the warm dust component in NGC\,4565, we find a spatial correspondence with the dust heated by young stellar objects in NGC\,4565 and the molecular gas reservoir, which provides the birth material for this star formation. The resemblance between the MIPS 24\,$\mu$m, SPIRE 250\,$\mu$m and IRAM 1.2 mm dust profiles suggests that the cold and warm dust component are well mixed in the interstellar medium of NGC\,4565. However, the asymmetry in CO observations from \citet{1994PASJ...46..147S} does become evident in the dust emission profiles of the warmer dust component. This asymmetry is most likely the consequence of spiral arm structures in the plane of the galaxy. Similar asymmetric absorption features could also be identified in optical images \citep{1980PASJ...32..197H,1996AJ....112..114R} and were assigned to a spiral arm structure at position angle $\sim$ 135$^{\circ}$.

\subsection{Other infrared/(sub)millimeter data}
\label{OtherIR}
To refine the sampling of the spectral energy distribution, we include photometric data, other than those obtained by \textit{Herschel} and \textit{Spitzer}, ranging from the infrared to millimeter wavelength domain. 
At near- and mid-infrared wavelengths, we complement our dataset of observations with flux measurements from the Wide-field Infrared Survey Explorer (WISE, \citealt{2010AJ....140.1868W}). 
Images belonging to the WISE all-sky survey at 3.4, 4.6, 12.1 and 22.2\,$\mu$m were retrieved from the NASA/IPAC Infrared Science Archive. To measure fluxes from those images, all stars and background sources were masked in the field of the galaxy.
Since the galaxy is located at the edge of the map, a determination of the background level from an annulus around the galaxy was impossible.
A background value was determined from averaging the flux over 100 apertures taken randomly within a 30$\arcmin$ field in this masked image.
The size of the apertures was adapted to the FWHM of the PSF in every waveband with a diameter of 4 $\times$ FWHM. The FHWM in the different WISE bands varies from 6.1$\arcsec$, 6.4$\arcsec$, 6.5$\arcsec$ at 3.4, 4.6, 12.1\,$\mu$m up to 12$\arcsec$ in the longest 22.2\,$\mu$m waveband \citep{2010AJ....140.1868W}. The mean background value was subtracted from the masked image.
From this masked and background-subtracted image, we determined fluxes in all wavebands from aperture photometry. With the stellar bulge being a prominent emission feature at near-infrared wavelengths and the mid-infrared images having a dominant contribution from PAHs and very small grains in the star-forming disk of NGC\,4565, the size of the aperture was adjusted to fit the size of the galaxy emitting in each waveband.
Fluxes of the stellar emission from WISE at 3.4 and 4.6\,$\mu$m are consistent with IRAC flux measurements at 3.6 and 4.5\,$\mu$m (see Table \ref{dataptn}). Also the WISE emission in the longer wavelength bands corresponds well with measurements by IRAS and MIPS at similar wavelengths (see Table \ref{dataptn}).

IRAS flux measurements at 12, 25, 60 and 100\,$\mu$m were adopted from the Catalog of IRAS observations of large optical galaxies \citep{1988ApJS...68...91R}. Also the Kuiper Space Airborne Observatory observed NGC\,4565 \citep{1992ApJ...394..104E} in wavebands centered at 100, 160 and 200\,$\mu$m (see Table \ref{dataptn}).
ISO observations at 170\,$\mu$m (32.5 $\pm$ 4.9 Jy, \citealt{2004A&A...422...39S}) were not considered here due to aperture effects (see also \citealt{2009ApJ...707.1809W}).
With \textit{Akari} measurements at 65, 90, 140 and 160\,$\mu$m (1.17 $\pm$ 0.12 Jy, 3.99 $\pm$ 0.22 Jy, 11.85 $\pm$ 1.07 Jy and 12.58 $\pm$ 2.39 Jy, respectively) being up to one order of magnitude lower in comparison to results from other airborne and space telescope facilities (IRAS, ISO, MIPS, KAO) in overlapping wavebands, we decided not to take into account \textit{Akari} flux measurements for the analysis in this paper.

At millimeter wavelengths, IRAM observations at 1.2\,mm constrain the emission from cold dust in those wavebands.
\citet{Alton} reported a flux density of 55 mJy when integrating along the major axis profile of the 1.2 mm emission for a 20$\arcsec$ beam. Assuming the 1.2 mm emission is spatially distributed similar to that at 250\,$\mu$m, we estimate a total flux density of 99 mJy at 1.2\,mm for NGC\,4565 (see Table \ref{dataptn}).

\subsection{GALEX and optical data (0.15-0.89 micron)}
\label{GALEXoptical.sec}
GALEX $FUV$/$NUV$ and optical SDSS $ugriz$ data for NGC\,4565 were recovered from the GALEX and SDSS archives and reduced following the instructions outlined in \citet{Cortese}.
Final maps were reduced to obtain pixel sizes of 1.5$\arcsec$ and 0.4$\arcsec$ for GALEX and SDSS observations, respectively. GALEX $FUV$ and $NUV$ channels are characterized by a FHWM of 4.2$\arcsec$ and 5.3$\arcsec$, respectively \citep{2007ApJS..173..682M}, whereas the shape of the PSF in SDSS maps depends on the seeing during sky exposures.
GALEX images for NGC\,4565 were obtained in a single run of 1693 s exposure time, with SDSS images covered in 54 s of observing time.
The first two panels on the top row of Figure \ref{Images_all.pdf} show the GALEX $FUV$ and $NUV$ images, respectively.
Both GALEX images clearly show the low level of star formation activity in the inner most regions in NGC\,4565. This lack of direct UV emission from young stars in the inner 80$\arcsec$ region in NGC\,4565 was already identified by \citet{1992ApJ...394..104E} and \citet{1995PASJ...47...17O}, relating this deficiency in star formation to a bar structure in the central regions.
 
Figure \ref{Images_all.pdf} presents the SDSS $u$, $g$, $r$, $i$ and $z$ band images in the last two top panels and first three panels on the second row. 
The main emission features in the optical SDSS images are the stellar disk and bulge. At shorter optical wavelengths (in particular $u$ band), a large fraction of the stellar light is extinguished along the line-of-sight by the prominent dust lane. Towards longer wavelengths, the dust obscuration diminishes and also the boxy shape of the bulge becomes more evident.
This boxy shaped bulge is thought to result from a bar in the inner region of NGC\,4565 viewed edge-on \citep{1981A&A....96..164C} and is therefore considered to be a pseudo-bulge.
A genuine bulge component is also present in NGC\,4565, but its spatial extent and intensity is inferior to the dominant emission from the pseudo-bulge \citep{2010ApJ...715L.176K}.

Both from GALEX and SDSS images, the warping structures on the southeastern and northwestern sides of the galaxy's disk can be discerned.
Especially at UV wavelengths, the warps are prominent emission features, indicative for the ongoing star formation in these tidally disrupted parts of the interstellar medium (ISM) in NGC\,4565.
Flux densities for the stellar emission in UV/optical wavebands were adopted from \citet{Cortese} for GALEX $FUV$/$NUV$ and SDSS $gri$ bands (see Table \ref{dataptn}). Fluxes for the remaining SDSS bands ($u$, $z$) were obtained in an homogeneous way from aperture photometry similar to the procedure for the other SDSS bands.

\section{Dust energy balance in NGC\,4565}
\label{Balance.sec}
\subsection{SKIRT and FitSKIRT} 
\label{SKIRT.sec}
SKIRT \citep{2003MNRAS.343.1081B,Baes2011} is a 3D Monte Carlo radiative transfer code designed to model the absorption, scattering and thermal re-emission of dust in a variety of environments: circumstellar disks \citep{2007BaltA..16..101V, Vidal2011}, clumpy tori around active galactic nuclei \citep{2012MNRAS.420.2756S} and different galaxy types \citep{2010A&A...518L..39B,2010A&A...518L..54D,2010MNRAS.403.2053G, 2010A&A...518L..45G}. 

FitSKIRT \citep{DeGeyter} is a fitting routine that combines the output of SKIRT with a genetic algorithm optimization library to obtain a best fitting model for the stellar and dust components in a galaxy. In this manner,  the best fitting parameters for a variety of models and stellar and dust geometries can be derived when providing an input image from which the stellar emission and dust obscuration effects can be modelled. 

\subsection{Radiative transfer modeling}
\label{RTModel.sec}
\subsubsection{Model 1}
\label{Model1}
A first model accounting for the observed optical properties of stars and dust in NGC\,4565 is obtained from the FitSKIRT fitting algorithm. FitSKIRT was applied to determine the best fitting model parameters using our SDSS $g$ band image (see Figure \ref{Images_all.pdf}, last top panel) as a reference image. We choose the $g$-band image for the fitting procedure, since the emission in this waveband is dominated by the old stellar population. At shorter wavelengths, the contribution from young stars increases rapidly. The $g$ band image is also optimal to perceive the attenuation effects of dust, which diminishes at longer wavelengths. Prior to the FitSKIRT fitting algorithm, the stellar warps on both ends of the galactic disk were masked in this image.

\begin{figure*} 
\centering \includegraphics[width=0.95\textwidth]{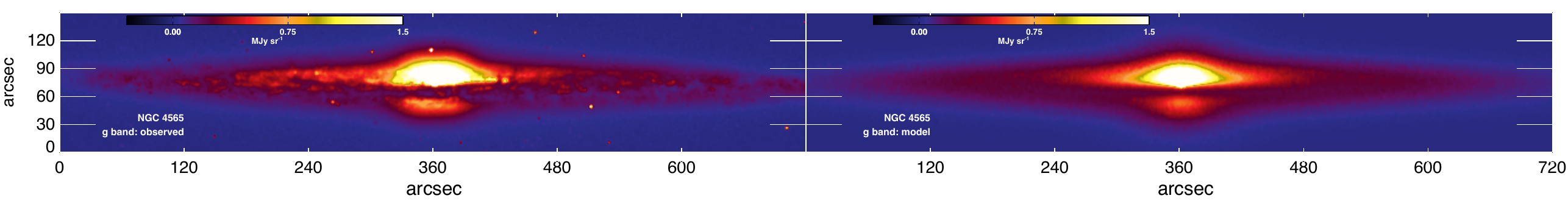}  \\ 
\includegraphics[width=0.24\textwidth]{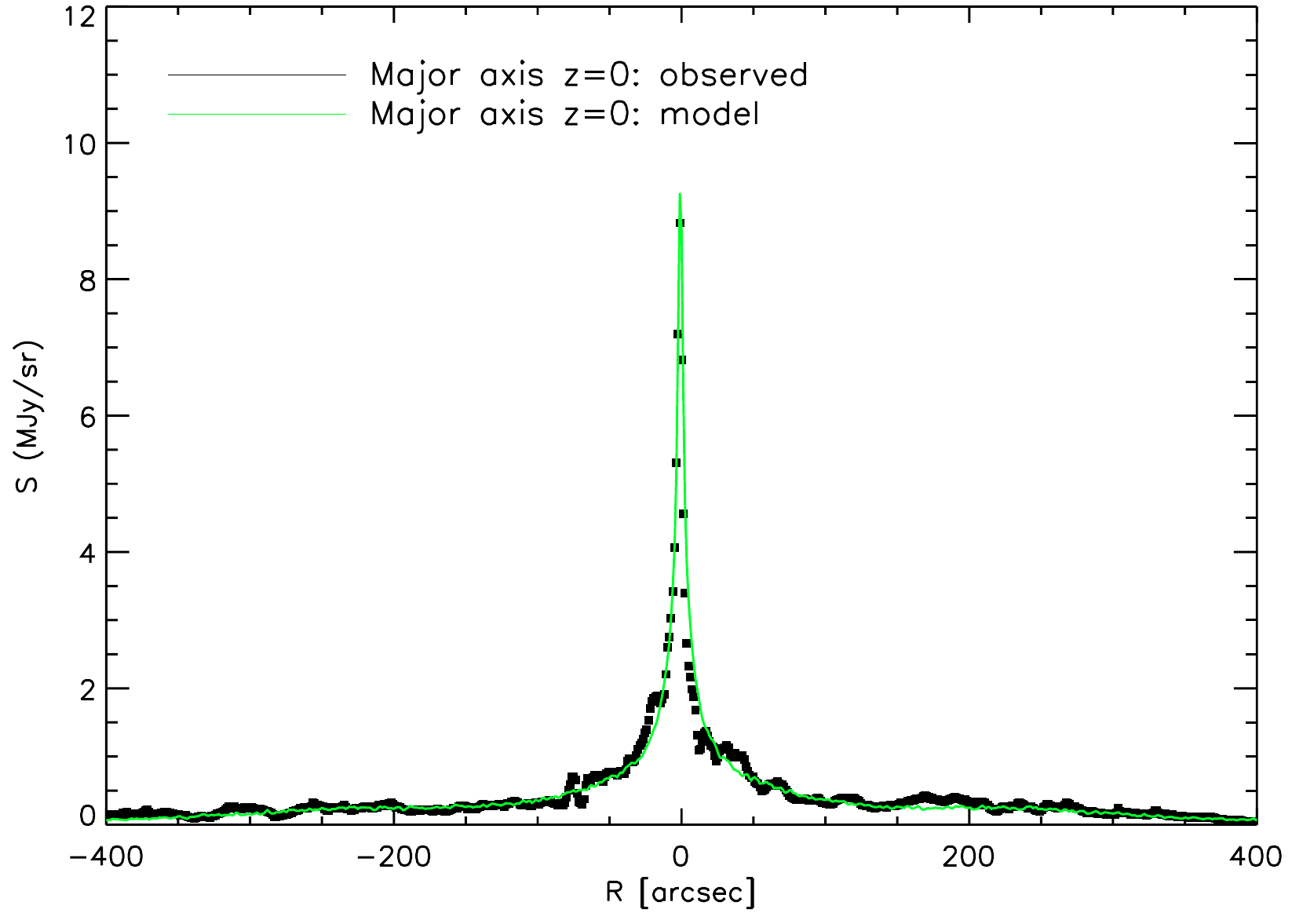} \includegraphics[width=0.24\textwidth]{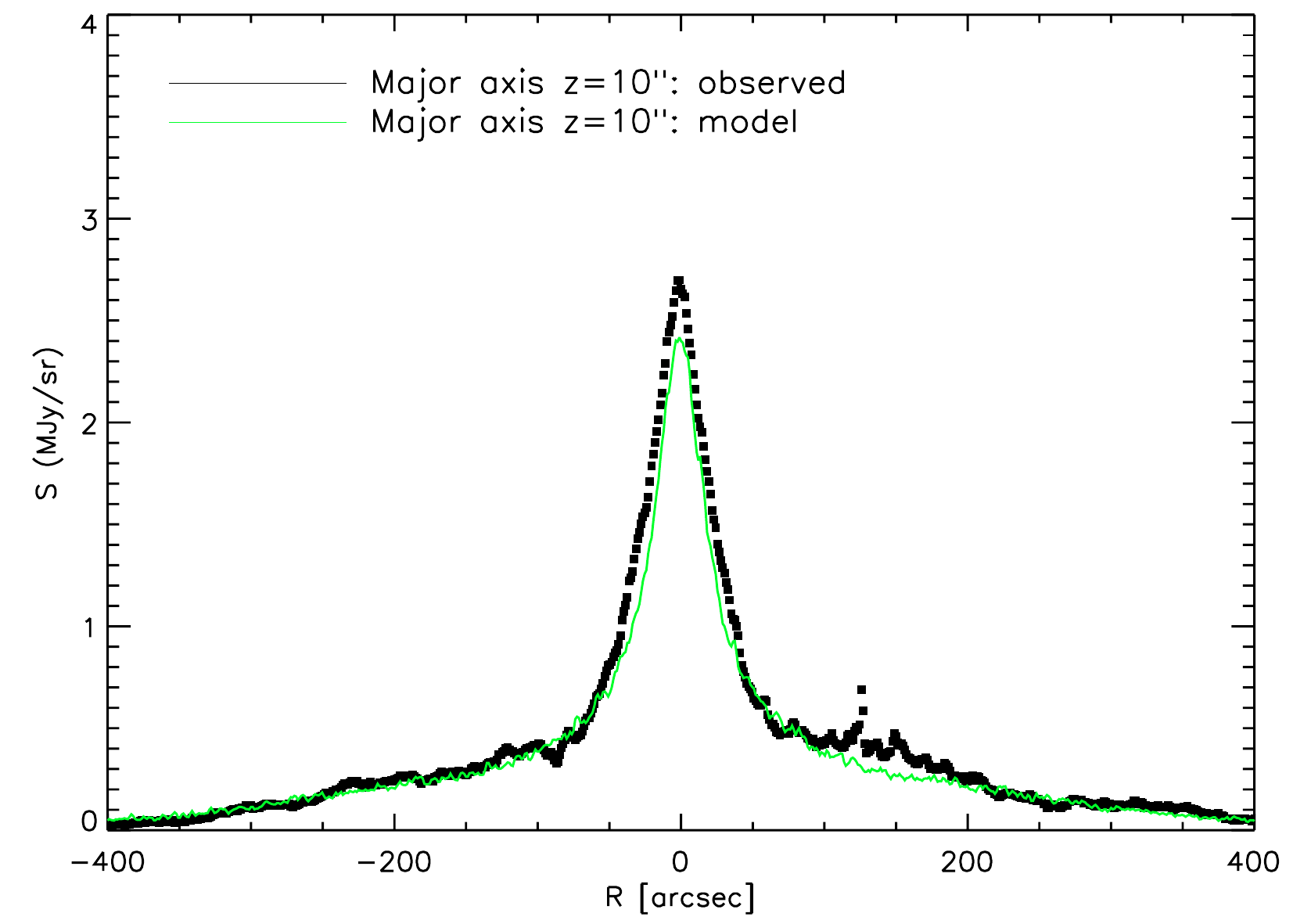}  
\includegraphics[width=0.24\textwidth]{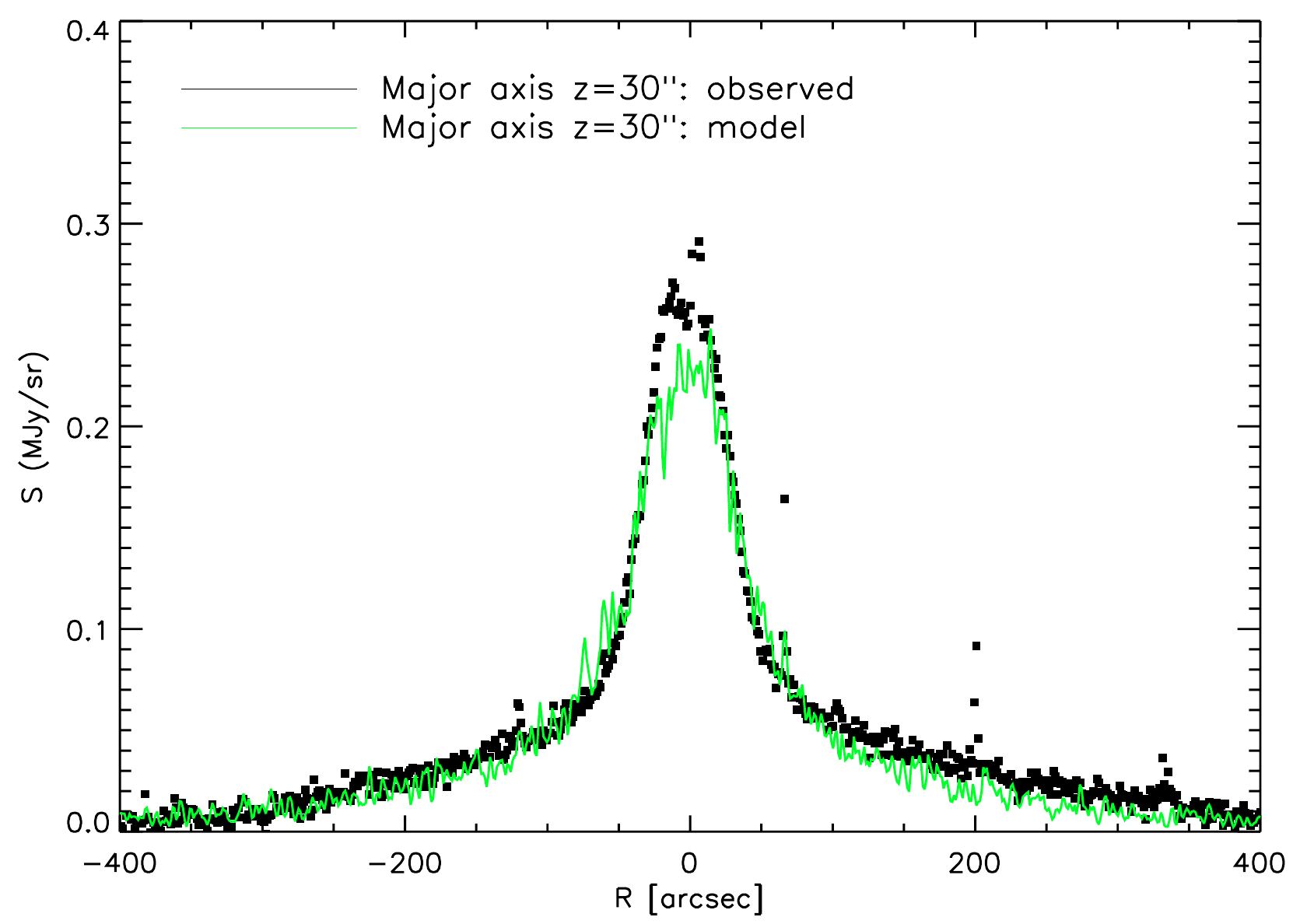}  
\includegraphics[width=0.24\textwidth]{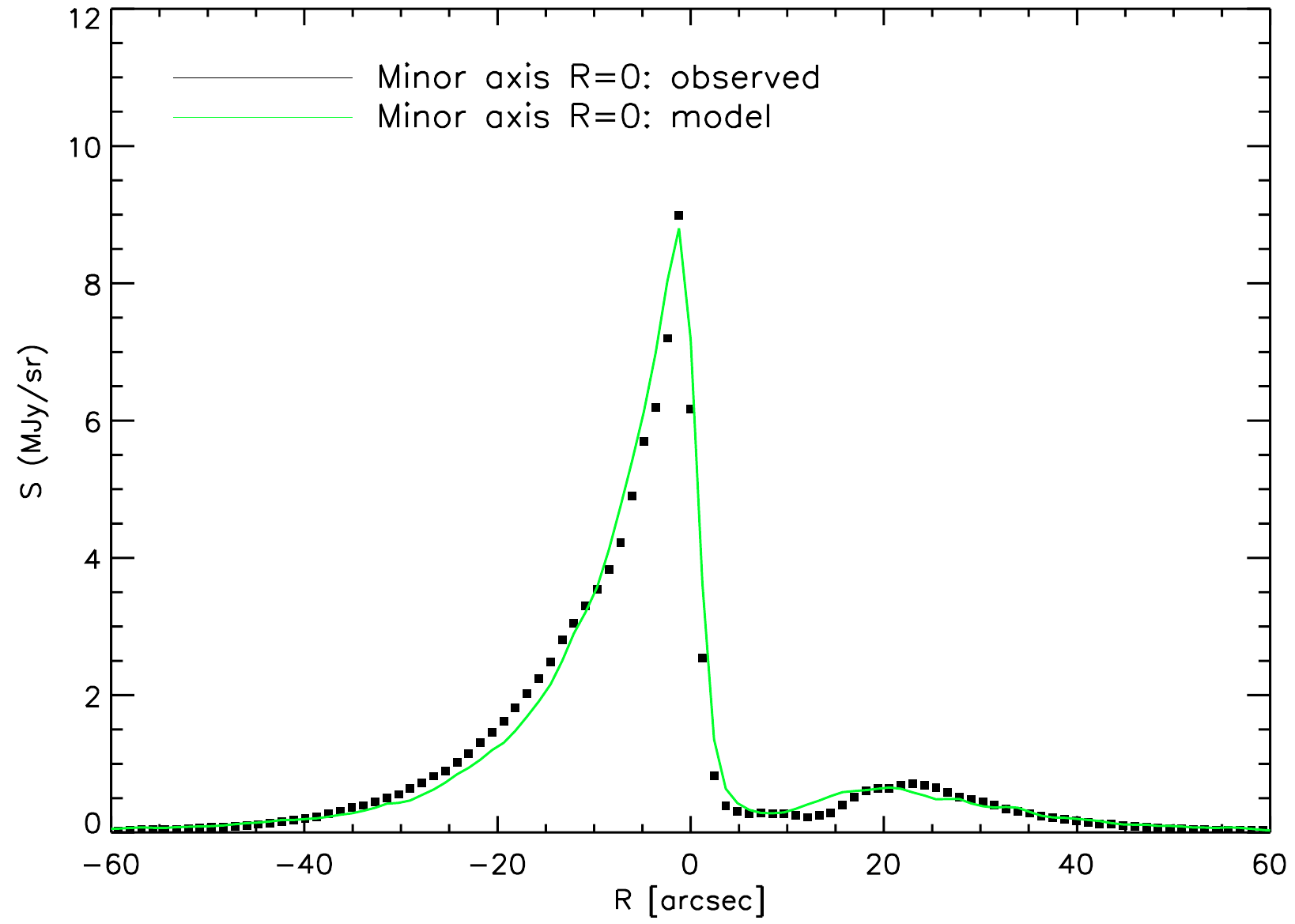} 
\caption{Model 1: the observed (left) and modelled (right) $g$ band image for NGC\,4565 on the top row. The bottom row shows from left to right the major axis $g$ band profiles at radial distances z=0, 10 and 30$\arcsec$ from the center of NGC\,4565, respectively. The rightmost panel represents the minor axis profiles at R=0. The observed profiles are indicated in black, with the modelled emission color-coded in green.}  \label{Ima_g_model1.pdf}
\end{figure*}

The input model for the stellar component consists of three components. The first component is an exponential disk with scale length $h_{R}$ and height $h_{z}$ to fit the disk component
\begin{equation}
\label{expdisk}
\rho(R,z) = \rho_{0} \, \exp\left(-\frac{R}{h_{R}} \right) \exp\left(-\frac{|z|}{h_{z}} \right).
\end{equation}
The second component is a flattened Sersic geometry for the outer bulge component. It is modelled as
\begin{equation}
\label{sersic}
\rho(r) = \rho_0\, {\cal{S}}_n \left(\frac{m}{r_{\text{eff}}}\right),
\end{equation} 
where $m$ = $\sqrt{R^2+\frac{z^2}{q^2}}$ is the spheroidal radius ($q$ is the intrinsic flattening or axial ratio) and ${\cal{S}}_n(s)$ represents the dimensionless 3D spatial density of a model that deprojects to a projected surface brightness profile
\begin{equation}
I(r_p) = I_0 \exp
    \left[ -b_n\left( \frac{r_p}{r_{\text{eff}}} \right)^{1/n}
    \right],
\end{equation}
where $b_n$ is a dimensionless constant that can be approximated by the formula: $b_{\text{n}}$~=~2n~-~$\frac{1}{3}$~+~$\frac{4}{405 n}$~+~$\frac{46}{25515 n^2}$~+~$\frac{131}{1148175 n^3}$, as derived by \citet{1999A&A...352..447C}.
Finally, the third component consists of an inner bulge described as a Sersic function (same as Equation \ref{sersic} but with $m$ = $\sqrt{R^2+ z^2}$) with a Sersic index $n$ $\sim$ $1.4$ and a scale length $r_{\text{eff}}$ $\sim$ 1.3$\arcsec$, similar to the values reported in \citet{2010ApJ...715L.176K}. The luminosity of the inner bulge was scaled to a fixed value $L_{\text{V}}$ $\sim$ 3.6 $\times$ 10$^{4}$ L$_{\text{V},\odot}$, optimized to reproduce the $g$ band minor-axis profile (see Figure \ref{Ima_g_model1.pdf}, last bottom panel). 

For the spectral energy distribution (SED) of the stars, we assume a \citet{1998MNRAS.300..872M, 2005MNRAS.362..799M} single stellar population (SSP) parametrization. The age and metallicity of the SSP in the disk and bulge are fixed to 8 Gyr old and a solar metallicity ($Z$ $=$ 0.02), as obtained from diagnostic plots of several Lick indices (see Figures 5 and 6 in \citealt{2000MNRAS.311...37P}).
The age of the SSP is assumed to be the same in the stellar disk and bulge, since the boxy bulge is considered to be a pseudo bulge.
The intensity of the stellar components is set by its $V$ band luminosity ($L_{\text{V}}$).

Earlier studies of the dust geometry in NGC\,4565 have suggested a ring-like structure for a large fraction of the dust \citep{1992ApJ...394..104E,1995PASJ...47...17O,1996A&A...310..725N,2010ApJ...715L.176K,2010AJ....140..753L}.
Therefore, our dust model is composed of an exponential disk with a density distribution as represented in Eq. \ref{expdisk} and a ring with a Gaussian distribution in radial direction and an exponential vertical profile
\begin{equation}
\label{ring}
\rho(R,z) = \rho_{0} \,
    \exp\left[ -\frac{(R-R_0)^2}{2\sigma^{2}} \right] \exp\left(
    -\frac{|z|}{h_z} \right),
\end{equation}
where $R_0$, $\sigma$ and $h_{z}$ describe the radius of the ring, the radial dispersion and the vertical scale height.
The content of the dust reservoir in NGC\,4565 is scaled by the dust mass in each of the dust components.
The dust population is assumed to be uniform across the entire galaxy, consisting of a composition of dust particles with a fixed grain size distribution. The abundances, extinction
and emissivity of the dust mixture are taken from the \citet{2007ApJ...657..810D} model for the dust in our own Galaxy.

With the different model components for the stars and dust in NGC\,4565 and furthermore assuming a variable inclination for the galaxy, the degrees of freedom in the fitting procedure amount to 15 parameters.  
Our model does not account for the continuum emission from the AGN in the center of NGC\,4565, since the low [Ne III] 15.55\,$\mu$m-to-[Ne II] 12.81\,$\mu$m ratio ($\sim$ 1, \citealt{2010AJ....140..753L}) suggest that the AGN continua are weak in this galaxy \citep{2007ApJ...671..124D}.
Table \ref{model4} gives an overview of the best fitting parameters obtained from the FitSKIRT algorithm. Since a sufficient number of photon packages were used to obtain this best fitting model (a photon package is a collection of a number of photons, with the ensemble of all photon packages representing the radiation field in the galaxy), we assume an uncertainty of about 20$\%$ for all derived parameters (see also \citealt{DeGeyter}).

Exploring the compatibility of this model with optical data, we compare the observed $g$ band image (see Figure \ref{Ima_g_model1.pdf}, top row) and the major- and minor-axis $g$ band profiles for several galacto-centric radii (see Figure \ref{Ima_g_model1.pdf}, bottom row) with the modelling results. The resemblance between modelled and observed images, major- and minor-axis profiles suggests that our radiative transfer model is representative for the observed stellar emission and optical properties of dust in NGC\,4565.
For the inclination, we find a best fitting value of $\sim$ 87.25$^{\circ}$, which is in agreement with earlier reported results.
Indeed, H{\sc{i}} observations of the gaseous disk in the past have constrained the inclination $>$ 84$^{\circ}$ \citep{1991Ap&SS.177..465R}, with more recent H{\sc{i}} observations reporting an inclination angle of 87.5$^{\circ}$ (Zschaechner et al. in prep.).
From a similar radiative transfer study, \citet{Alton} estimated an inclination of 88$^{\circ}$.

From the stellar distribution in this radiative transfer model, we infer a bulge-to-disk ratio of B-to-D $\sim$ 0.47 in the $V$ band. 
The model indicates a total dust mass of $M_{d}$ $\sim$ 1.0 $\times$ 10$^{8}$ M$_{\odot}$ of which about 60 $\%$ resides in an exponential disk, with the remaining 40 $\%$ distributed in a broad ring structure. The scale size of the exponential dust disk ($h_{R}$ $\sim$ 223$\arcsec$, $h_{z}$ $\sim$ 2.5$\arcsec$)  agrees well with the results obtained from a similar radiative transfer modelling procedure in \citet{Alton} ($h_{R}$ $\sim$ 200$\arcsec$, $h_{z}$ $\sim$ 4.3$\arcsec$). The diffuse dust component in \citet{Alton} has a face-on optical depth $\tau_{V}^{f}$ $\sim$ 0.63, which corresponds to $M_{d}$ $\sim$ 9.7 $\times$ 10$^{7}$ M$_{\odot}$, when assuming the same extinction cross-section for the \citet{2007ApJ...657..810D} dust composition applied in our dust model.

\begin{figure*} 
\centering 
\includegraphics[width=0.80\textwidth]{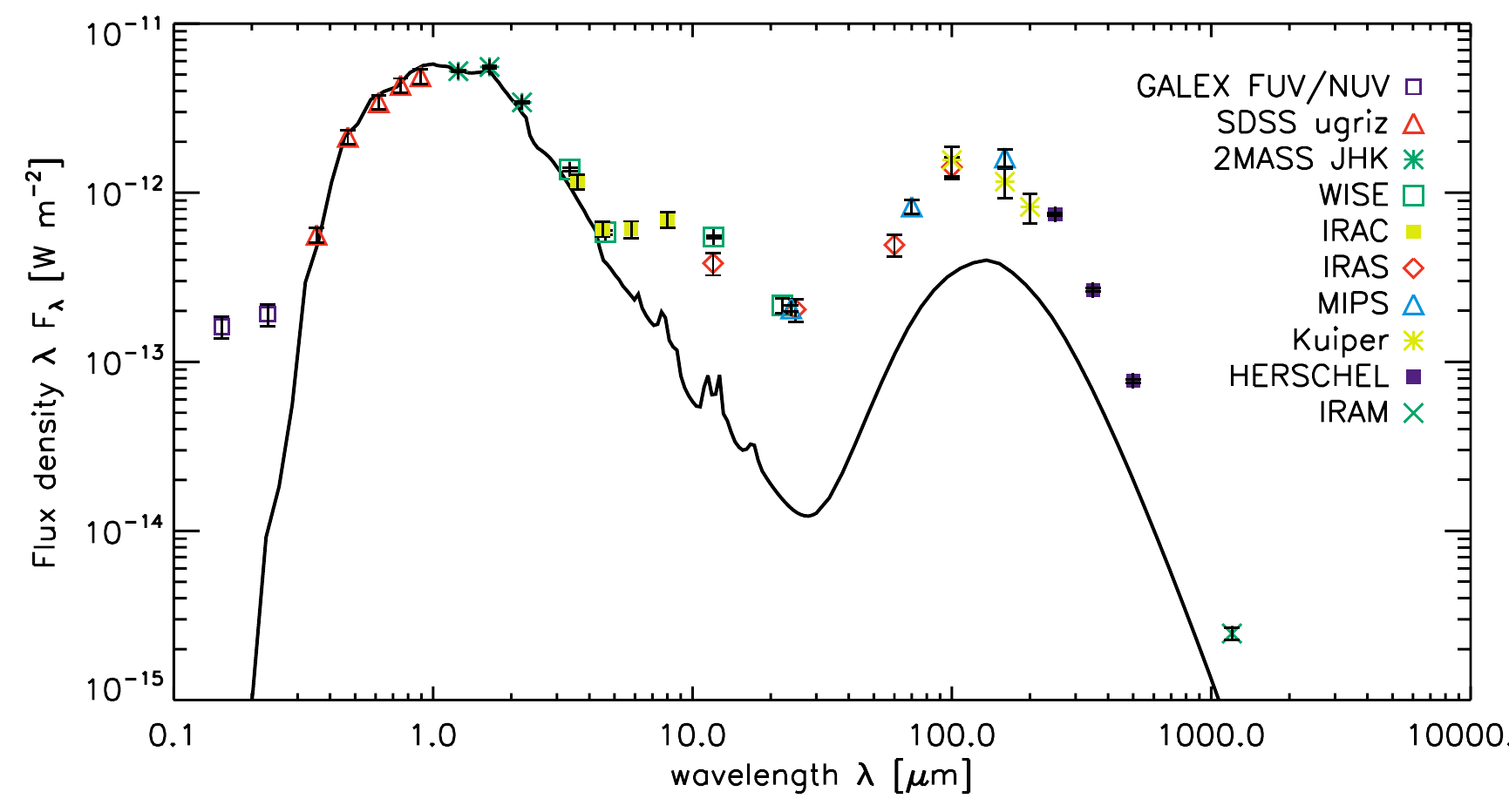} 
\caption{Model 1: the modelled SED as obtained with the SKIRT model (solid black line), consisting of merely an old stellar population and neglecting the contribution from young stars and/or a clumpy dust distribution, overlaid with the observed fluxes (see Table \ref{dataptn}).}  \label{SED_model1.pdf}
\end{figure*}

Figure \ref{SED_model1.pdf} represents the spectral energy distribution as modelled by our radiative transfer code SKIRT (solid, black line). Available flux measurements were overlaid on this SED model to describe the observed spectral energy distribution. An overview of flux measurements for NGC\,4565 is given in Table \ref{dataptn}. 
Less accurate flux measurements were considered non-relevant for this analysis and were omitted from this plot (see Section \ref{Obs.sec}).
Although our model shows great resemblance with the $g$ band image and major- and minor axis profiles, the spectral energy distribution in Figure \ref{SED_model1.pdf} shows that our model fails to reproduce the emission from NGC\,4565 in the UV and MIR/FIR wavelength domains.
The higher observed UV and MIR emission from NGC\,4565 indicates the presence of a young stellar population in this galaxy, which is not yet accounted for by our model. 

\subsubsection{Model 2: standard model with star formation}
\label{Model2}
In an attempt to reproduce the ultraviolet and infrared emission observed from NGC\,4565, we add a star formation component to the model constructed in Section \ref{Model1}.
Rather than using an empirical star formation template (e.g. \citealt{2011ApJ...741....6M}), we adopt starburst templates from the library of pan-spectral SED models for the emission from young star clusters with ages $<$ 10 Myr presented in \citet{2008ApJS..176..438G}. The SED templates in \citet{2008ApJS..176..438G} were generated from a one-dimensional dynamical evolution model of H{\sc{ii}} regions around massive clusters of young stars, the stellar spectral synthesis code Starburst 99 \citep{1999ApJS..123....3L} and the nebular modelling code MAPPINGS III \citep{Groves}. The Starburst99 templates used in these models correspond to an instantaneous burst with a \citet{2002Sci...295...82K} broken power-law IMF. The main parameters controlling the shape of the emission spectrum from these young stellar clusters are the metallicity ($Z$) of the gas, mean cluster mass ($M_{\text{cl}}$), age and compactness ($C$) of the stellar clusters, the pressure of the surrounding ISM ($P_{0}$) and the cloud covering fraction ($f_{\text{PDR}}$). 
The metallicity is chosen similar to the metallicity of the old stellar population in NGC\,4565 ($Z$ $\sim$ 1 Z$_{\odot}$). To eliminate the age parameter, we use age-averaged templates obtained from averaging spectra over 21 cluster ages, ranging from 0.01 to 10 Myr in steps of 0.5 Myr. For the mean cluster mass and ISM pressure, we assume fixed values $M_{\text{cl}}$ $\sim$ 10$^5$ M$_{\odot}$ and $P_{0}$/$k$ $\sim$ 10$^{6}$ cm$^{-3}$ K. Those approximations are justified since more massive star clusters can be simply thought of as the superposition of several individual clusters.  A variation in the ISM pressure furthermore mainly affects the nebular emission lines rather than altering the shape of the emission spectrum \citep{2008ApJS..176..438G}. For the cloud covering fraction $f_{\text{PDR}}$, we will assume a value of $f_{PDR}$~=~1 to describe the emission from heavily obscured star-forming regions in NGC\,4565. 
The choices for $M_{\text{cl}}$ and $P_{0}$/$k$ leave us to explore the effect of the compactness parameter $C$ on the shape of the SED for values of log $C$ ranging between 4.5 and 6.5. With the compactness parameter controlling the position of the far-infrared emission bump by shifting it in peak wavelength without significantly altering the width of the emission curve, the parameter $C$ is closely related to the dust temperature \citep{2008ApJS..176..438G}. Although the emission at 24\,$\mu$m might still be dominated by hot dust emission predominantly heated by young stars of ages $<$ 10 Myr, the contribution of diffuse dust emission is non-negligible at far-infrared and sub-millimeter wavelengths. 
Therefore, it is difficult to constrain the shape of the emission spectrum from hot dust grains and thus to determine the appropriate value of log $C$ in the models. 
With values of 6.5 for log $C$ describing extremely compact star-forming regions, which are characteristic for the high pressure ISM and/or massive young clusters in starburst galaxies, we assume a more moderate compactness factor of log $C$~=~5.5.
A value of log $C$ = 5.5 is furthermore consistent with the dust temperature $T_{d}$~=~30 K found for the warm dust component heated predominantly by stellar emission originating from star-forming regions \citep{1992ApJ...394..104E}.

In addition to these young stellar clusters, we have also included a component of ultra-compact H{\sc{ii}} regions in our model. Those ultra-compact H{\sc{ii}} regions are typically used to simulate the conditions during the earliest stages of the cluster lifetime ($<$ 10$^{6}$ yr) when the newly formed stars are likely still buried in their separate birth clouds and a typical star cluster consists of an ensemble of individual ultra-compact H{\sc{ii}} regions, rather than providing a single source of radiation originating from the smoothed emission from all stars in the stellar cluster. The templates for these ultra-compact H{\sc{ii}} regions were constructed in a similar way as for the dusty H{\sc{ii}} regions with their surrounding PDRs, but each birth cloud is modelled individually and the template for the entire compact H{\sc{ii}} regions is constructed by co-adding the individual SEDs (see also \citealt{2008ApJS..176..438G} for a more thorough description of the models). 
The SED template for the ultra-compact H{\sc{ii}} is jointed with the SED template for the H{\sc{ii}} regions with $f_{PDR}$~=~1, under the assumption that they produce new stars at the same rate.
This joint SED template will be scaled until the observed emission at 24\,$\mu$m is reproduced by our model.

Emission spectra for the young stellar population with ages between 10 and 100 Myr are taken from the Starburst99 library \citep{1999ApJS..123....3L}.
The Starburst99 templates were constructed for an instantaneous burst with a solar metallicity, a burst mass $M_{cl}$~=~10$^{6}$ M$_{\odot}$ and a \citet{2002Sci...295...82K} broken power-law IMF, which is consistent with the stellar emission spectra used to compute the SED of stellar clusters younger than 10 Myr in \citet{2008ApJS..176..438G}. With the young stellar clusters $<$ 10 Myr emitting predominantly in mid-infrared wavebands, the non-ionizing radiation of the evolved stellar population up to ages 50-100 Myr will dominate the UV emission spectrum in NGC\,4565 \citep{2005ApJ...633..871C}.

\begin{figure}[th!] 
\centering 
\includegraphics[width=0.45\textwidth]{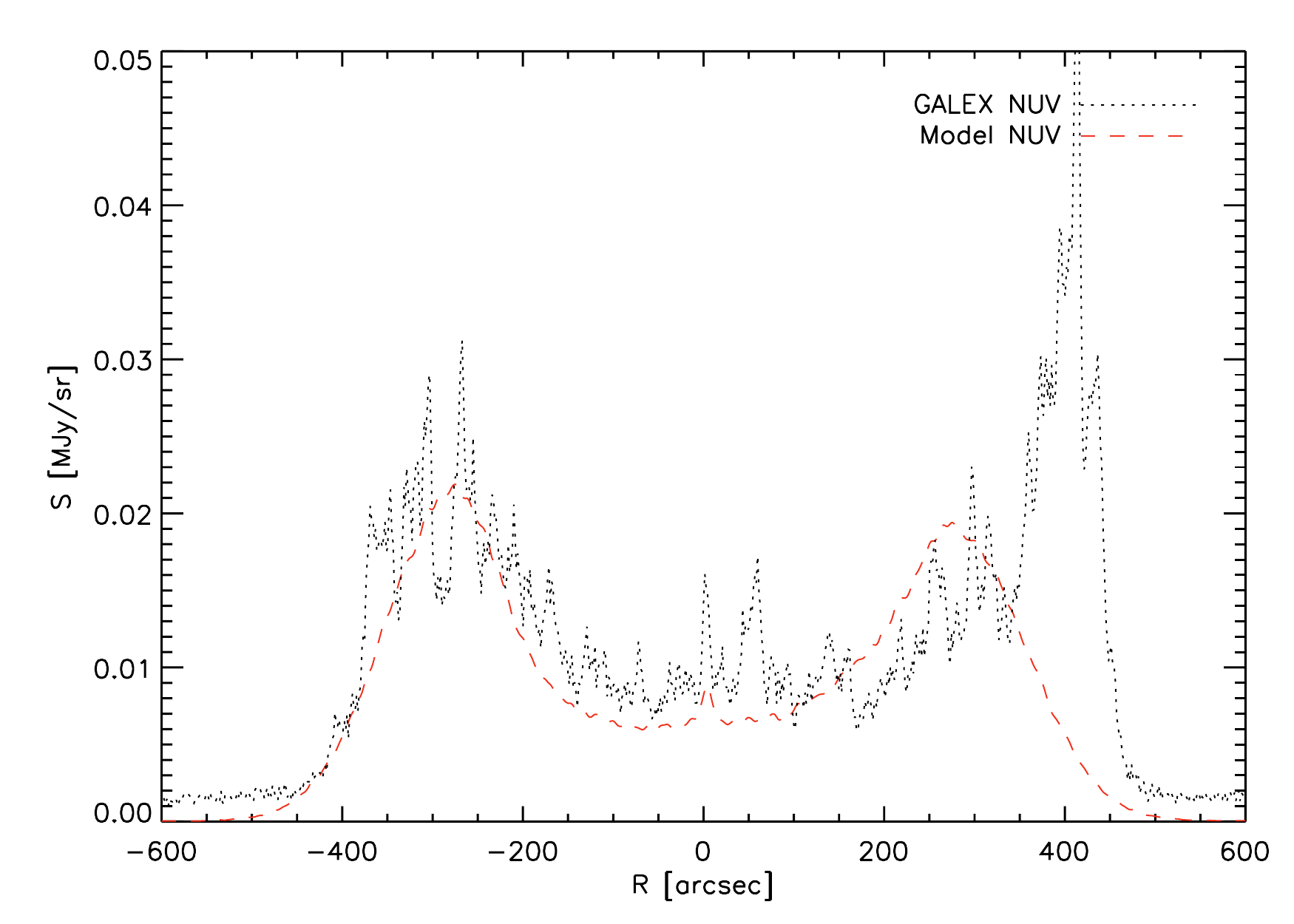} \\
\includegraphics[width=0.45\textwidth]{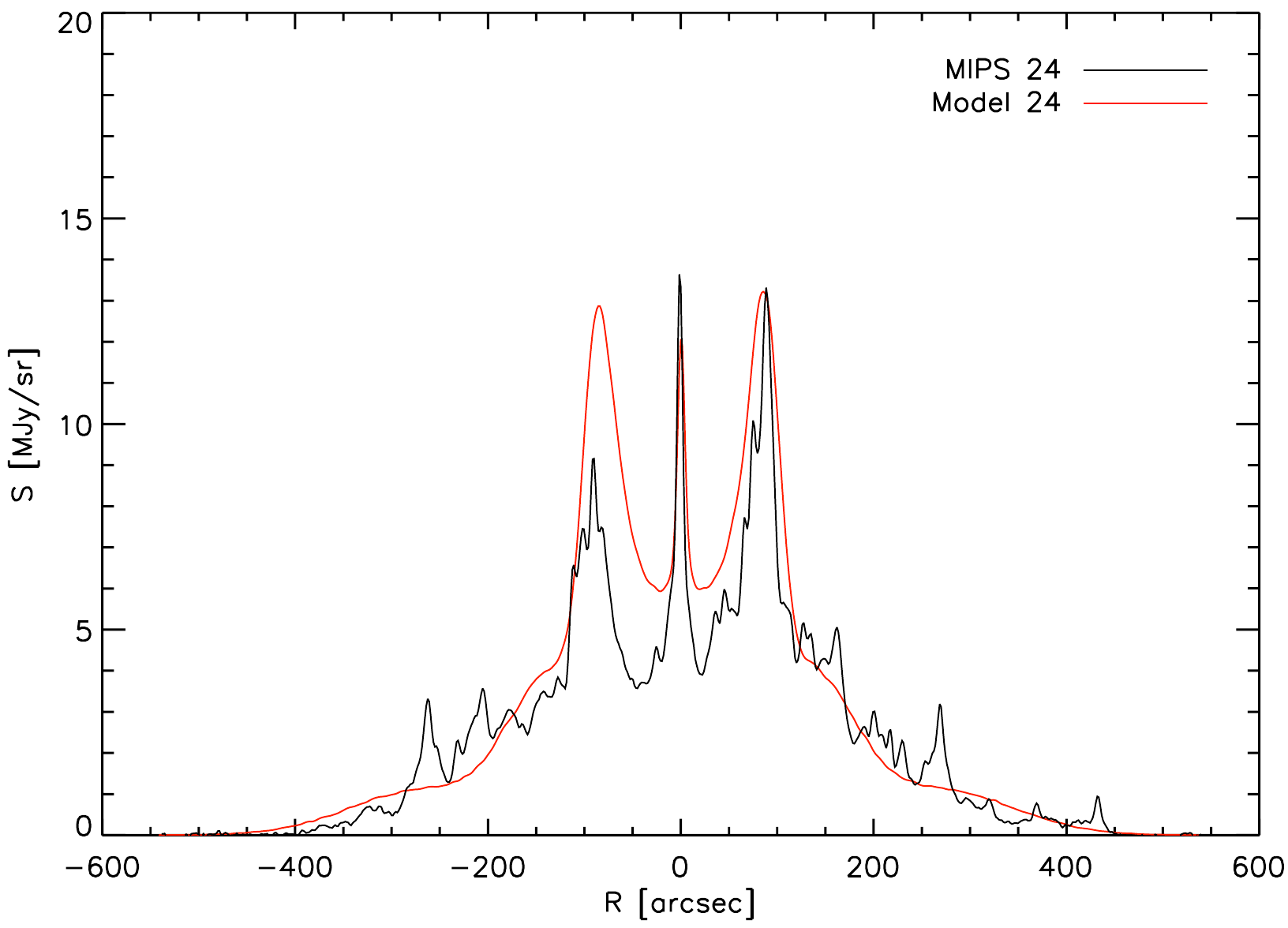} 
\caption{Model 2: $NUV$ (top) and MIPS 24\,$\mu$m (bottom) major axis profiles. The observed profiles are indicated in black, with the modelled emission color-coded in red.}  \label{Major_UV_MIPS24.pdf}
\end{figure}

Although most details on the exact location of star-forming complexes within the disk vanish along the line of sight, we try to recover the average geometry of unobscured and dust-enshrouded star-forming regions from the emission profiles along the major axis of NGC\,4565 in the GALEX $NUV$ and MIPS 24\,$\mu$m wavebands (see Figure \ref{Major_UV_MIPS24.pdf}, top and bottom panel, respectively).
By allocating a certain distribution to the star-forming regions, the proper amount of attenuation due to the foreground screen of diffuse dust is applied on every specific location within the galaxy.
Indeed, due to the high inclination angle the different star-forming regions in NGC\,4565 will experience wide ranges in obscuration depending on whether they are located on the near- or far end of the galaxy's disk with respect to the line-of-sight.
From the major axis $NUV$ emission profile, we derive a ring-like geometry for the unobscured star formation component, modelled as a ring (see Equation \ref{ring}) with central radius $R$ $\sim$ 290$\arcsec$ and width $\sigma$ $\sim$ 72$\arcsec$.
The scale height of the ring of young stars is assumed to be identical to the scale height of the stellar disk populated with the old stellar population ($h_{z}$ $\sim$ 8.2$\arcsec$).
From the MIPS 24\,$\mu$m image, it becomes immediately evident that a component of obscured star formation resides in a narrow ring coinciding with the CO molecular ring centered at 80-100$\arcsec$ \citep{1994PASJ...46..147S,1996A&A...310..725N}. 
The parameters for the ring (see Equation \ref{ring}) of embedded star formation are chosen similar to the molecular CO ring with central radius at $R$ $\sim$ 90$\arcsec$, width $\sigma$ $\sim$ 3.4$\arcsec$ and scale height $h_{z}$ $\sim$ 1.5$\arcsec$. 
Aside from this concentration of embedded localized sources distributed in a ring, a more extended component of obscured star formation seems also present in NGC\,4565. To reproduce the more extended 24\,$\mu$m emission in the major axis profile, obscured star-forming complexes are distributed in the ring harboring the unobscured star formation component and the dust ring containing $\sim$ 40$\%$ of the diffuse dust component in NGC\,4565. 
From the 24\,$\mu$m major axis profile (see Figure \ref{Major_UV_MIPS24.pdf}, bottom panel), the presence of a central dust disk becomes evident. 
This central dust concentration is modelled as an exponential disk with radial and vertical scale lengths of 5$\arcsec$ and 1$\arcsec$, respectively. The distribution of this central dust reservoir ($M_{d}$ $\sim$ 3 $\times$ 10$^4$ M$_{\odot}$) is truncated at a radius of 20$\arcsec$. 

When fitting the UV and 24\,$\mu$m emission of NGC\,4565, we require an average star formation rate in the outer regions of the disk (with most of the young stars between galacto-centric radii of 200$\arcsec$ and 400$\arcsec$) of 3.1 M$_{\odot}$ yr$^{-1}$ for the 10 to 100 Myr old stellar population.
The heavily obscured star-forming regions of ages $<$ 10 Myr seem to reproduce the observed 24\,$\mu$m emission when assuming an average star formation rate of 2.2 M$_{\odot}$ yr$^{-1}$ over the past 10 Myr.  
The dust-enshrouded star-forming clumps in the narrow ring produce young stars at a rate of SFR $\sim$ 0.6 $M_{\odot}$ yr$^{-1}$, with a somewhat higher productivity for the dense clouds distributed in the ring of unobscured star formation (SFR $\sim$ 0.8 $M_{\odot}$ yr$^{-1}$) and the diffuse dust ring (SFR $\sim$ 0.8 $M_{\odot}$ yr$^{-1}$).
This obscured star formation rate inferred from the models is perfectly consistent with the SFR $\sim$ 2.0 M$_{\odot}$ yr$^{-1}$ calculated in \citet{2006PASP..118.1098W} from the far-infrared luminosity $L_{\text{FIR}}$ in NGC\,4565, corrected for the contribution from an old stellar population \citep{2005ApJ...633...86S}. Also monochromatic SFR tracers such as the 24\,$\mu$m emission result in similar estimates for the SFR $\sim$ 2.0 M$_{\odot}$ yr$^{-1}$ \citep{2009ApJ...692..556R}. We argue furthermore that the star formation rate of stars between 10 and 100 Myr in our model (SFR $\sim$ 3.1 M$_{\odot}$ yr$^{-1}$) is in close agreement with the SFR $\sim$ 2.7 M$_{\odot}$ yr$^{-1}$ calculated from the FUV luminosity $L_{\text{FUV}}$ for NGC\,4565 and relying on the SFR calibration in \citet{Salim}. 

Accounting for all young stellar objects in our model, we derive an average star formation rate of 5.3 M$_{\odot}$ yr$^{-1}$ over the past 100 Myr. The small decrease in the star formation activity during the recent 10 Myr is in agreement with a delayed exponential law describing the star formation history in other late-type spiral galaxies \citep{2001AJ....121..753B,2002ApJ...576..135G}. The total star formation activity in our model (SFR $\sim$ 5.3 M$_{\odot}$ yr$^{-1}$) is furthermore consistent with the total star formation rates derived from a combination of unobscured and obscured star formation tracers (e.g. SFR$_{\text{FUV+24}\mu \text{m}}$ $\sim$ 3.2 M$_{\odot}$ yr$^{-1}$ \citep{2008ApJ...686..155Z}, SFR$_{\text{FUV+TIR}}$ $\sim$ 3.9 M$_{\odot}$ yr$^{-1}$ \citep{2005ApJ...619L..51B}).

\begin{figure*}
\centering 
\includegraphics[width=0.98\textwidth]{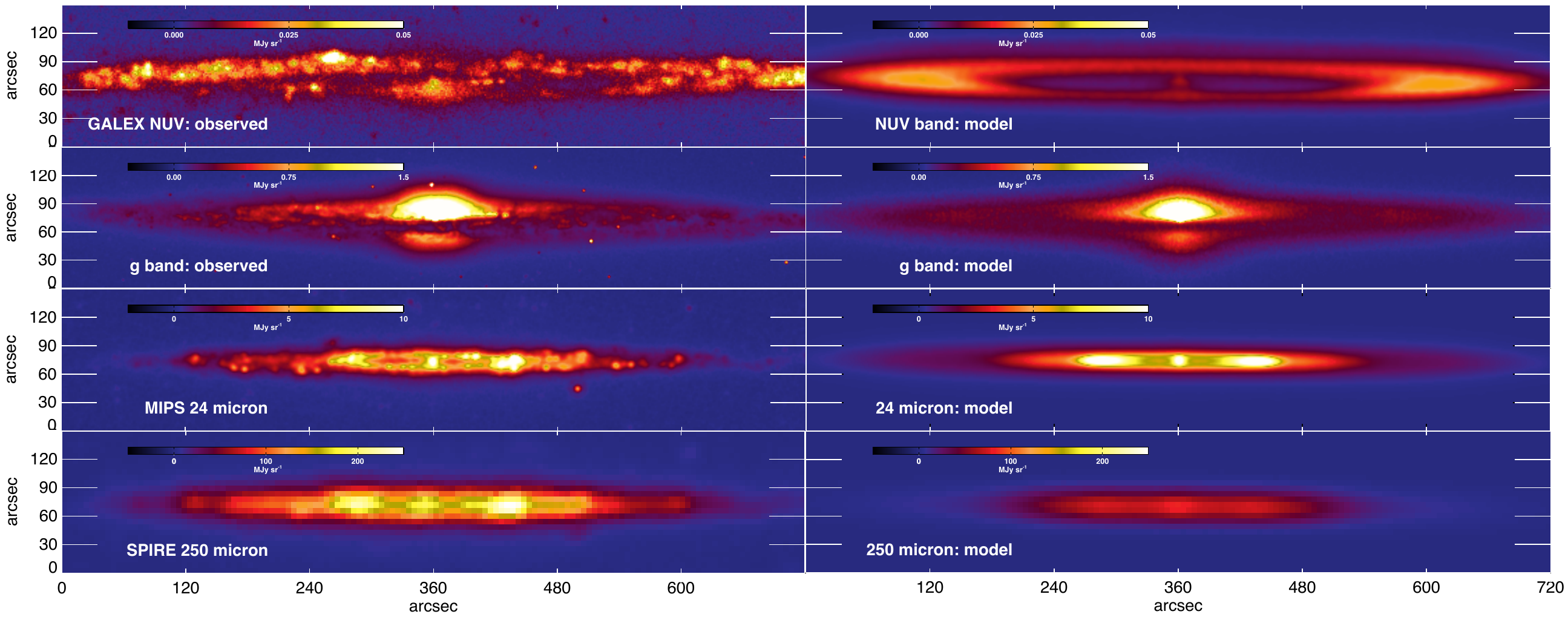} \\
\includegraphics[width=0.24\textwidth]{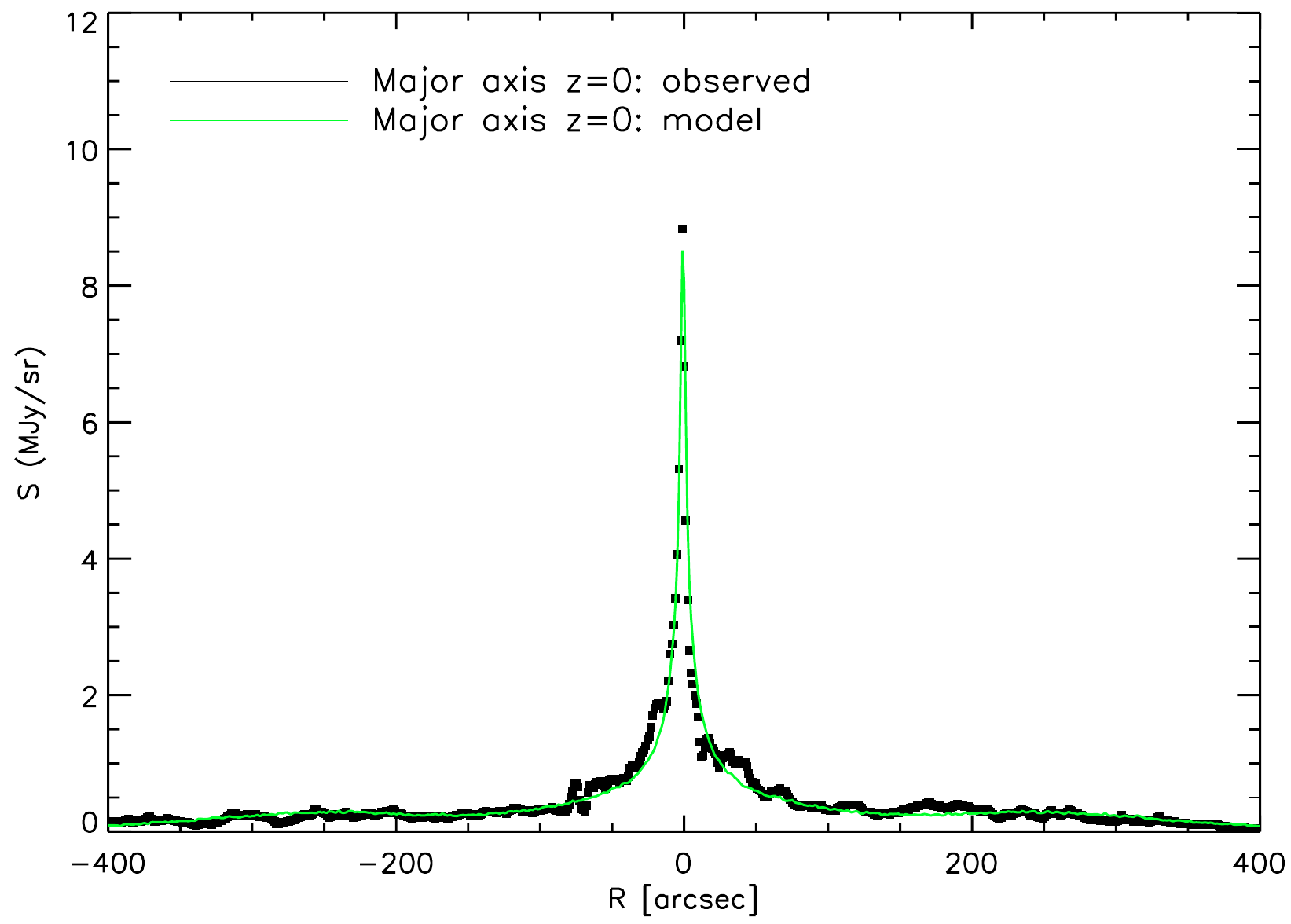} \includegraphics[width=0.24\textwidth]{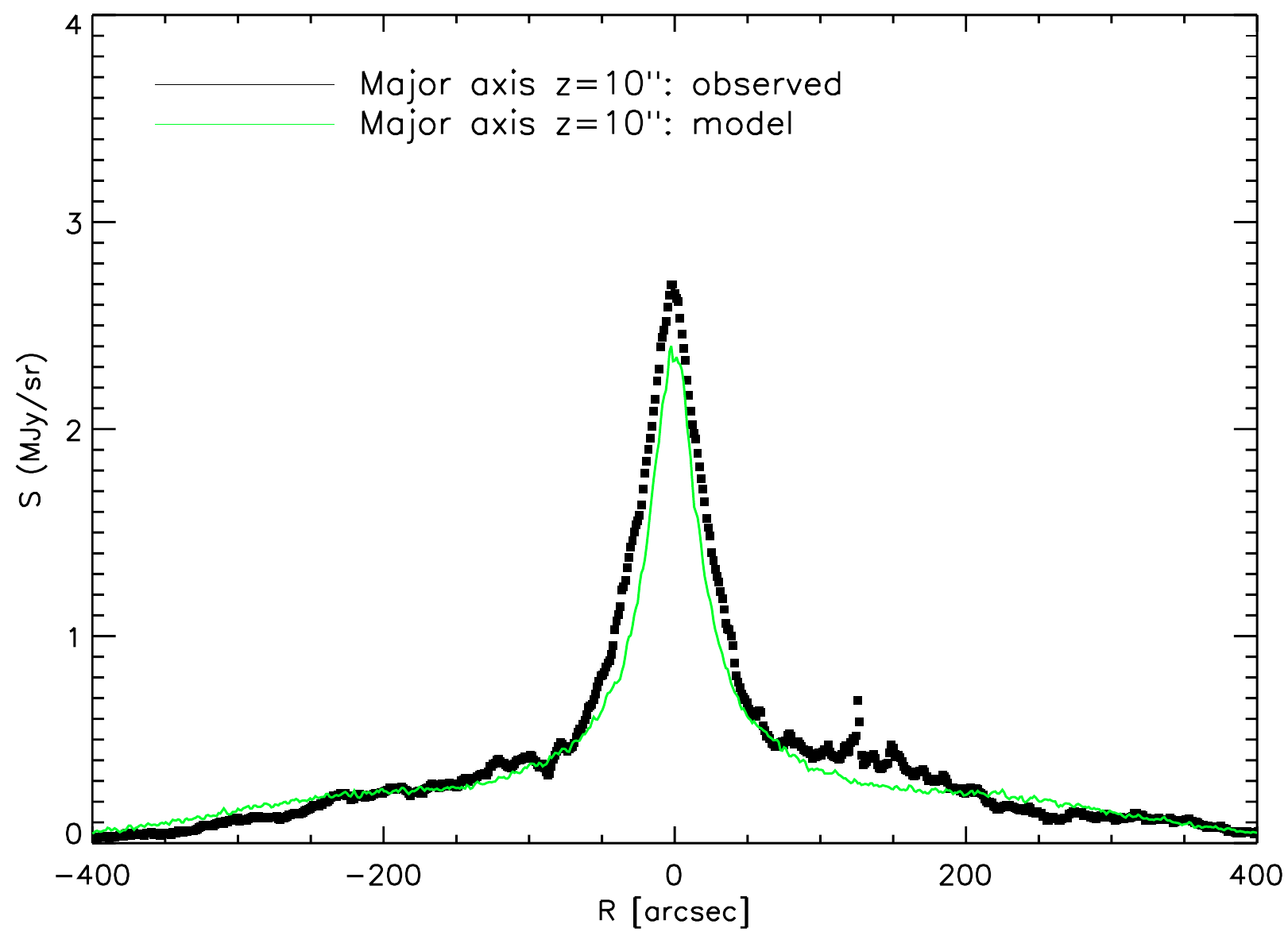}  
\includegraphics[width=0.24\textwidth]{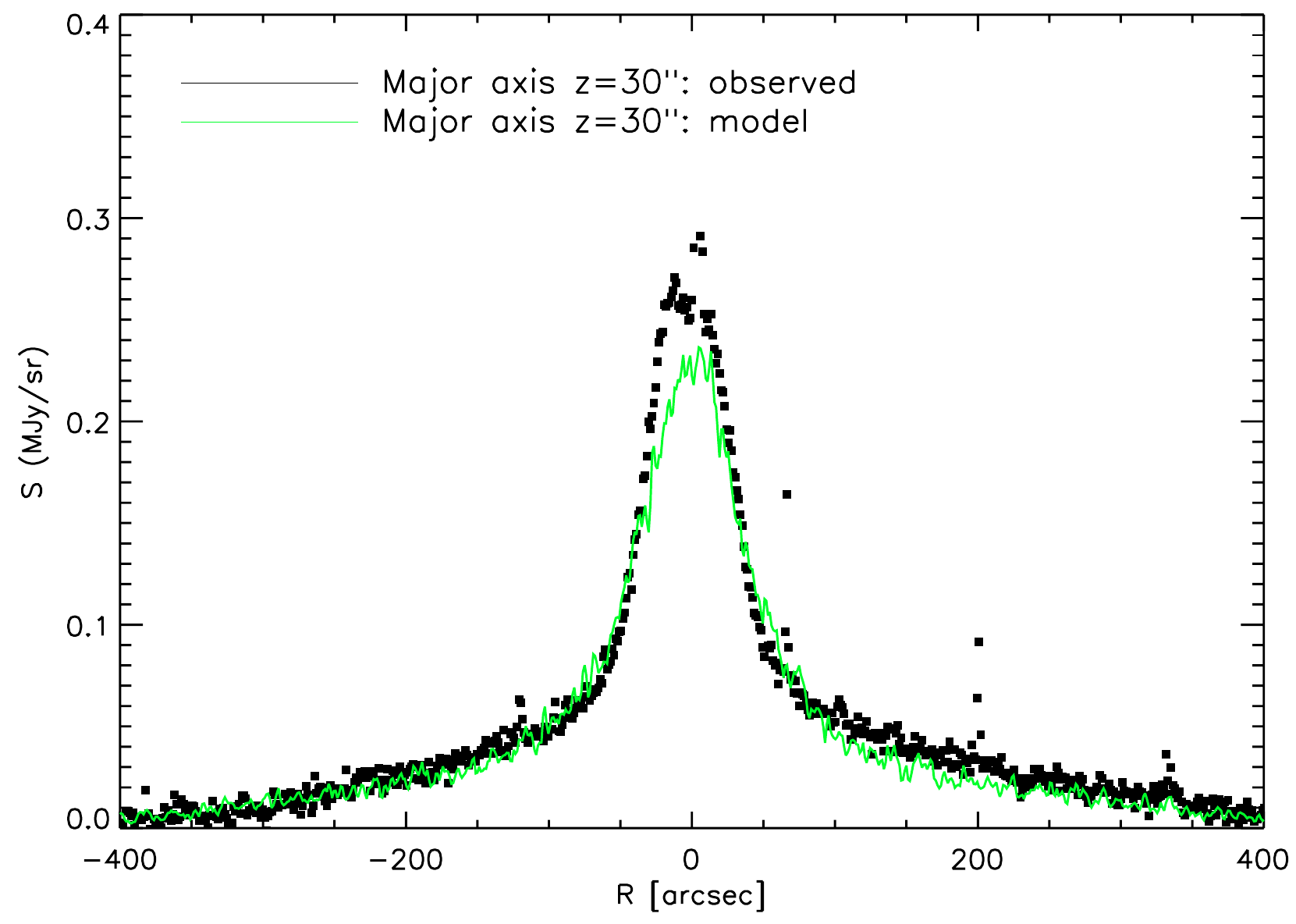}  
\includegraphics[width=0.24\textwidth]{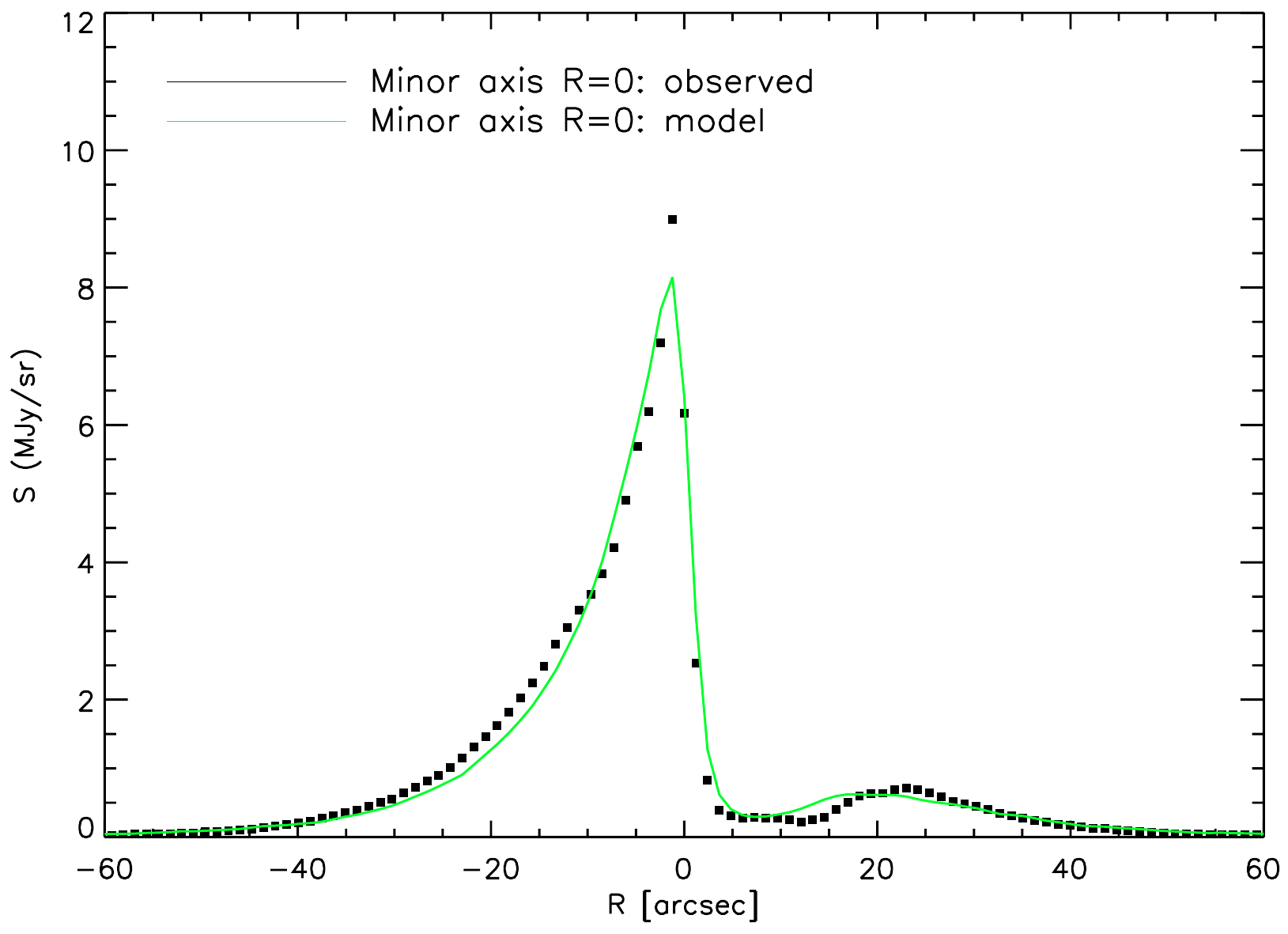} \\
\caption{Model 2 with on the top row: the observed (left) and modelled (right) $NUV$ band (top row), $g$ band (second row), 24\,$\mu$m (third row) and 250\,$\mu$m image (fourth row). The bottom row represents from left to right the major axis $g$ band profiles at radial distances z=0, 10 and 30$\arcsec$ from the center of NGC\,4565, respectively. The rightmost panel represents the minor axis profiles at R=0. The observed profiles are indicated in black, with the modelled emission color-coded in green.}  \label{Ima_model2.pdf}
\end{figure*}

This model with both unobscured and obscured star formation is able to account for the young stellar emission at UV/mid-IR wavelengths (see images on the first and third row of Figure \ref{Ima_model2.pdf}, respectively) and remains in agreement with the optical constraints (see images on the second row and plots on the last row of Figure \ref{Ima_model2.pdf}) upon reducing the luminosity of the old stellar population in the disk by a small factor to correct for the emission from those young stars in optical wavebands ($L_{\text{V}}$ = 4.0 $\times$ 10$^{10}$ L$_{\odot,\text{V}}$). The total dust mass associated with the young stellar clusters ($<$ 10 Myr) in our models is $M_{d}$ 3.6 $\times$ 10$^{7}$ M$_{\odot}$, which increases the total dust mass in NGC\,4565 to $M_{d}$ 1.4 $\times$ 10$^{8}$ M$_{\odot}$.
The geometry for the embedded, localized sources in NGC\,4565 is able to account for the 24\,$\mu$m emission in NGC\,4565 as observed from the major axis profile (see Figure \ref{Major_UV_MIPS24.pdf}, bottom panel) or the MIPS 24\,$\mu$m image (see images on the fourth row of Figure \ref{Ima_model2.pdf}).
The geometry for the less obscured young stars with ages $>$ 10 Myr is capable of reproducing the emission profile along the major axis (see Figure \ref{Major_UV_MIPS24.pdf}, top panel) as well as the emission originating from the entire galaxy at NUV wavelengths (see Figure \ref{Ima_model2.pdf}, third row). The peak at $R$ $\sim$ 420$\arcsec$ in the NUV major axis profile belongs to the residual star formation in the warp structure on the northwestern side of the galaxy's disk and is not taken into account in our model. The stellar clusters do have a more clumpy distribution throughout the disk than was accounted for by our model.

\begin{figure*}[th!] 
\centering 
\includegraphics[width=0.80\textwidth]{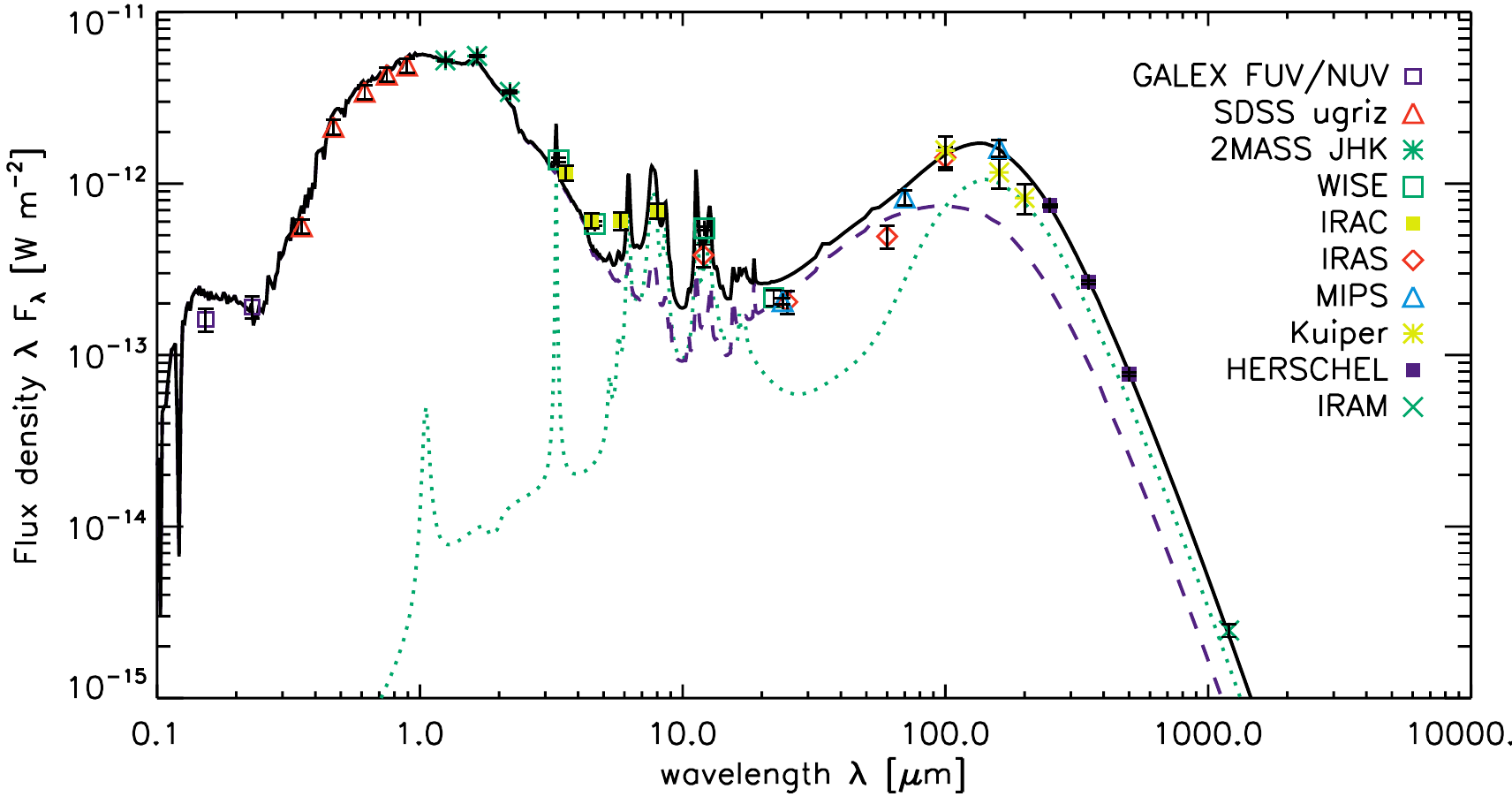}
\caption{Model 2: the modelled SED as obtained with the SKIRT model (dashed purple line) overlaid with the observed fluxes (see Table \ref{dataptn}).
Model 2 corresponds to model 1 supplemented with young star clusters ($<$ 100 Myr). 
To estimate the dust mass residing in quiescent dust clumps in NGC\,4565, we fit the DustEm dust model with a \citet{2007ApJ...657..810D} grain composition (green, dotted line) to the residual fluxes for NGC\,4565 in the far-infrared and submillimeter wavebands.
The model with star formation and an additional dust reservoir distributed in clumps can explain the multi-wavelength observations for NGC\,4565 (see black, solid curve).}
\label{SED_model2bis.pdf}
\end{figure*}

More critical is the inconsistency in the FIR/submm, where our radiative transfer model continues to underestimate the observed emission spectrum for NGC\,4565 (see Figure \ref{SED_model2bis.pdf}). 
Even though the model accounts for the dust heating provided by star formation, the modelled dust emission underestimates the observed far-infrared and submillimeter emission by a factor of $\sim$ 3-4, in correspondence to dust energy balance studies for other nearby edge-on spirals.

At mid-IR wavelengths, our model also underestimates the observed emission originating mainly from PAHs in those wavebands by a factor of $\sim$ 2 (see Figure \ref{SED_model2bis.pdf}). This dissimilarity might on the one hand be a consequence of the fixed PAH template used in the models to construct the emission spectra reported in \citet{2008ApJS..176..438G}. Based on the average variation by a factor of 2 obtained from a study of the mid-IR emission for a large sample of nearby galaxies \citep{2007ApJ...656..770S}, we should interpret this off-set as a diagnostic for the specific ISM conditions in NGC\,4565 rather than attributing it to a deficiency of star formation in our model.
With the PAH template in \citet{2008ApJS..176..438G} being optimized to fit the Spitzer IRS observations of two interacting galaxy pairs (NGC\,4676 and NGC\,7252), small deviations in ISM conditions and interstellar radiation field hardness might explain the difference in observed PAH emission for NGC\,4565. 
On the other hand, the diffuse emission from cold dust at FIR/submm wavelengths has not yet properly been accounted for by our radiative transfer model and might be responsible for an important fraction of the PAH emission in galaxies (e.g. \citealt{2002A&A...385L..23H, 2008MNRAS.389..629B}).

\begin{table}
\caption{Best fitting model parameters for the stellar (exponential disk~+~outer bulge) and dust components (exponential disk~+~ring) in NGC\,4565 as obtained from FitSKIRT.}
\label{model4}
\begin{center}
\begin{tabular}{@{}|clc|}
\hline
Geometry & $i$ [$^{\circ}$] & 87.25  \\
\hline \hline 
Stellar disk & $h_{\text{R}}$~[$\arcsec$] & 102 \\
 & $h_{\text{z}}$~[$\arcsec$] & 8.2 \\
 & $L_{\text{V}}$ [L$_{\odot,\text{V}}$]\footnotemark[1] & 4.9 $\times$ 10$^{10}$   \\
\hline \hline 
Outer bulge & $n$ & 2.12 \\
 & $r_{\text{eff}}$~[$\arcsec$] & 20.2 \\
 & $q$ & 0.744 \\
 & $L_{\text{V}}$ [L$_{\odot,\text{V}}$] & 2.0 $\times$ 10$^{10}$ \\
\hline \hline 
Dust disk & $h_{\text{R}}$~[$\arcsec$] & 223 \\
 & $h_{\text{z}}$~[$\arcsec$] & 2.5 \\
 & $M_{d}$~[M$_{\odot}$] & 6.3 $\times$ 10$^7$ \\
\hline \hline 
Dust ring &  $R$~[$\arcsec$] & 146 \\
 & $h_{\text{z}}$~[$\arcsec$] & 4.3 \\
 & $\sigma$~[$\arcsec$] & 41.2\\
 & $M_{d}$~[M$_{\odot}$] & 4.0 $\times$ 10$^7$ \\
\hline
\end{tabular}
\end{center}
\footnotemark[1]{The luminosity of the stellar disk is later adjusted to $L_{\text{V}}$ = 4.0 $\times$ 10$^{10}$ L$_{\odot,\text{V}}$ to correct for the emission from young stars in the $V$ band when supplementing our model with star formation.} 
\end{table}

\section{Discussion}
\label{Discussion.sec}
\subsection{Dust energy balance}
Although in agreement with the observed properties of stars and dust at optical wavelengths and the young stellar emission in UV and MIR wavebands, the model from section \ref{Model2} underestimates the observed dust emission in the far-infrared and (sub)millimeter wavelength regime. Whereas the dust emissivity of dust grains could influence the shape of the SED at some wavelengths \citep{Alton}, the inconsistency in the integrated dust energy balance can not be solely caused by a grain population with dust emissivities diverging from Galactic values at longer wavelengths. Indeed, the multi-frequency sampling of the dust SED allows us derive a true discrepancy in the integrated dust energy balance, which is impossible to solve by assuming divergent FIR emissivities due to a different grain composition and structure compared to the typical Galactic dust reservoir.
Although we can not exclude a possible variation in dust emissivity at submm/mm wavelengths (diverging from the assumed emissivity in the \citealt{2007ApJ...657..810D} dust model), we argue that this discrepancy is likely due to a clumpy morphology of the dust in NGC\,4565 unresolved in the currently available dataset of far-infrared/submillimeter observations for NGC\,4565. Due to the high density in those clumps, they hardly contribute to the attenuation of stellar light and can therefore not be distinguished from optical observations.

Due to the edge-on inclination of NGC\,4565 and the current limitations in resolving power of infrared instrumentation, we are not capable of constraining the exact location and size of those dust clumps and rather have to remain content with information on the average mass and temperature of this clumpy dust reservoir. 
Those dust clumps either correspond to quiescent clouds where star formation has not (yet) been initialized. On the other hand, this extra dust reservoir might reside in the outer shells of star-forming complexes where the heating by the young embedded objects becomes negligible due to the high optical depth of the dust shells in the immediate vicinity of the stellar source.

To estimate the amount of dust residing in clumps, we fit the DustEm dust model with a \citet{2007ApJ...657..810D} grain composition to the residual fluxes after subtracting the dust emission in our model from the observed flux densities (IRAC 5.8, 8.0\,$\mu$m, WISE 12.1\,$\mu$m, IRAS 12 and 100\,$\mu$m, Kuiper 100 and 160\,$\mu$m, MIPS 160\,$\mu$m, SPIRE 250, 350, 500\,$\mu$m, IRAM 1.2\,mm).
Figure \ref{SED_model2bis.pdf} shows the resulting SED (black, solid curve) when supplementing our model with a dust reservoir of $\sim$ 1.5 $\times$ 10$^{8}$ $M_{\odot}$ at a temperature of $T_{d}$ $\sim$ 16.9 K (green, dotted curve). 
The average temperature $T_{d}$ $\sim$ 16.9 K obtained for the residual dust reservoir is consistent with the temperature of the diffuse dust in the disk of NGC\,4565 ($T_{d}$ $\sim$ 14-18 K) and the average colour temperatures derived in other nearby spiral galaxies at wavelengths longwards of 160\,$\mu$m (e.g. M81, M83 and NGC\,2403, \citealt{2012MNRAS.419.1833B}).
At mid-IR wavelengths, this additional dust reservoir is capable of explaining the PAH emission in NGC\,4565. Also at FIR/submm wavelengths, a massive cold dust component seems adequate to explain the emission in those wavebands. At 60 and 70\,$\mu$m, our radiative transfer model however overestimates the observed fluxes. This inconsistency could be an indication for the star-forming complexes in NGC\,4565 to be more compact than currently accounted for by our model. Due to the high inclination angle of NGC\,4565, a direct identification of heavily obscured star-forming complexes within the disk of the galaxy and their individual size distribution and compactness is impeded.

With these additional cold dust clumps, the total dust mass in NGC\,4565 amounts to $M_{d}$ $\sim$ 2.9 $\times$ 10$^{8}$ $M_{\odot}$ of which one-third has a diffuse distribution throughout the disk and the remaining two-thirds resides in clumps. From those dense dust clouds, the majority ($\sim$ 80$\%$) is not heated by UV-emitting sources while the remainder ($\sim$ 20$\%$) host embedded star-forming complexes.
The total dust mass $M_{d}$ $\sim$ 2.9 $\times$ 10$^{8}$ M$_{\odot}$ in our model is consistent with the dust reservoir accounted for by BLAST observations \citep{2009ApJ...707.1809W}, but $\sim$ 10 times higher than the dust content reported in \citet{1992ApJ...394..104E} (2.7 $\times$ 10$^{7}$ M$_{\odot}$) based on IRAS observations up to 100\,$\mu$m. This confirms the need for longer wavelength observations to trace the entire dust reservoir in a galaxy (e.g. \citealt{2010A&A...518L..65B,2010A&A...518L..89G,2012MNRAS.419.1833B}).

The solution to the dust energy budget problem for NGC\,4565 accounting for a sizable fraction of the dust ($\sim$ 65$\%$) distributed in compact clumps is in agreement with other energy balance studies of edge-on objects (e.g. \citealt{2008A&A...490..461B,2010A&A...518L..39B,2011A&A...527A.109P,2012MNRAS.419..895D,2012arXiv1204.2936H}, etc.).  

\subsection{Gas-to-dust ratio}
From the H{\sc{i}} mass $M_{\text{HI}}$ $\sim$ 1.8 $\times$ 10$^{10}$ M$_{\odot}$ (Zschaechner et al. in prep.) and the CO observations ($M_{\text{H}_{2}}$ $\sim$ 2.9 $\times$ 10$^{9}$ M$_{\odot}$) presented in \citet{1996A&A...310..725N} (both quantities are already converted to our adopted distance of $D$ = 16.9 Mpc), we derive a total gas mass of $\sim$ 2.9 $\times$ 10$^{10}$ M$_{\odot}$ for NGC\,4565 when applying a correction factor of $\sim$ 1.4 to correct for heavier elements.
Based on the total dust mass in our radiative transfer model ($M_{d}$ $\sim$ 2.9 $\times$ 10$^{8}$ M$_{\odot}$), we estimate an average gas-to-dust ratio of $\sim$ 100 in NGC\,4565, which is in fair agreement with the gas-to-dust fraction found in other late-type spirals (e.g. $\sim$ 120 in our own Galaxy, \citealt{2004ApJS..152..211Z}) and the dust-to-H{\sc{i}} mass fraction derived for Sb galaxies with normal H{\sc{i}} components (as opposed to H{\sc{i}} deficient objects) in the HRS sample of galaxies \citep{2012arXiv1201.2762C}.

\subsection{Dust heating sources}
Investigating the dust heating in NGC\,4565, we find that about 70$\%$ of the far-infrared luminosity is attributed to the absorption of stellar light from the old stellar population, with the heating for the remaining 30$\%$ provided by the young stellar population. In this calculation, we assume that the high-density dust clumps distributed throughout the diffuse ISM in NGC\,4565 are merely heated by the old stellar population, since the temperature of the dust ($T_{d}$ $\sim$ 16.9 K) in those dense dust cores does not suggest any link to localized embedded sources and is more closely related to the average temperature of the diffuse dust in the disk of the galaxy ($T_{d}$ $\sim$ 14-18 K). Those heating fractions are in agreement with similar studies of dust heating mechanisms in other nearby galaxies, where the old stellar population also accounts for an important fraction of the far-infrared radiation (e.g. \citealt{2008A&A...490..461B, 2010A&A...518L..65B, 2011AJ....142..111B, 2011A&A...527A.109P,2012MNRAS.419.1833B}). Similar results were obtained from accurate modelling of the absorption of stellar light as well as careful investigations of UV observations \citep{1996A&A...306...61B,2000ApJ...539..718C,2000ApJ...528..799W,2008MNRAS.386.1157C}.
\citet{1992ApJ...394..104E} report a dust heating efficiency of 54 and 46$\%$ for the young and old stellar population in NGC\,4565, respectively.
Considering that their IRAS observations only covered a wavelength range from 12 to 100\,$\mu$m, it is not surprising that they underestimated the contribution from the old stellar population to the dust heating with the more evolved stars contributing the bulk of dust heating at longer wavelengths.

Although the fraction of dust heated by the young stellar population in NGC\,4565 is overall small ($\sim$ 30 $\%$), the contribution of hot dust emission from star-forming complexes is about 90$\%$ in the 24\,$\mu$m waveband, after which it decreases quickly to about 50$\%$ at 70\,$\mu$m, 12$\%$ at 160\,$\mu$m, less than 10$\%$ in all SPIRE wavebands and below 5$\%$ at 1.2 mm.
Those values are consistent with the decreasing contribution of star-forming regions for increasing wavelength in M81, M83 and NGC\,2403 \citep{2012MNRAS.419.1833B}, with a relative contribution of the young stellar population to the total dust heating of more than 50$\%$ shortwards of 250\,$\mu$m and dropping below 30$\%$ in the SPIRE wavebands.

This self-consistent analysis of the dust heating mechanisms in NGC\,4565 based on a realistic radiative transfer model confirms earlier studies reporting on a non-negligible fraction of the dust heating in spiral galaxies powered by the more evolved stellar population.
With the young stellar population being responsible for less than half of the total infrared dust emission originating from NGC\,4565, we believe caution is needed when using the total infrared emission in galaxies to trace the star-forming activity. Although a tight correlation was found between the star formation and the total infrared dust emission for large samples of nearby spiral galaxies \citep{1990ApJ...350L..25D,1995AJ....110.1115D,1996A&A...306...61B,2009ApJ...703.1672K}, 
this relation might be the result of an indirect link between the star-forming activity in galaxies and the total-infrared emission in galaxies, tracing the surface density of gas in the interstellar medium. In the latter case, the SFR-$L_{FIR}$ correlation is governed by the Schmidt law \citep{1959ApJ...129..243S,1998ApJ...498..541K}, relating the surface density of gas and star formation in galaxies, rather than the dust heating provided by star-forming regions.   

\section{Conclusions}
\label{Conclusions.sec}

We present a full radiative transfer analysis of the edge-on spiral galaxy NGC\,4565, accounting for the absorption and scattering of stellar light by dust grains and its thermal re-emission at infrared wavelengths.
From a radiative transfer model fitting procedure to the optical SDSS $g$ band image, we determine the best fitting parameters for the old stellar population and diffuse dust component in NGC\,4565.
To account for the observed UV and mid-infrared emission in NGC\,4565, we supplement our model with a young stellar population of age $<$ 10 Myr (SFR $\sim$ 2.2 M$_{\odot}$ yr$^{-1}$) and ranging from 10 to 100 Myr (SFR $\sim$ 3.1 M$_{\odot}$ yr$^{-1}$). The distribution of star-forming complexes within the disk of the galaxy is constrained from major axis NUV and 24\,$\mu$m emission profiles.

Even though this young stellar population provides an additional power source for dust heating and hereby boost the emission at mid-infrared wavelengths, the emission observed at wavelengths longwards of 100\,$\mu$m remains underestimates by a factor of 3-4. This inconsistency in the dust energy budget of NGC\,4565 suggests the presence of a sizable fraction (two-thirds) of the total dust reservoir ($M_{d}$ $\sim$ 2.9 $\times$ 10$^{8}$ $M_{\odot}$) in a clumpy distribution with no associated young stellar sources. 
The distribution of those dense dust clouds would be such that they remain unresolved in current far-infrared/submillimeter observations and hardly contribute to the attenuation by dust at optical wavelengths. The contribution from a grain composition with dust emissivity properties diverging from the typical Galactic standards can however not be discarded as a factor of influence on the high far-infrared/submillimeter fluxes. 
The majority of dust heating in NGC\,4565 is provided by the old stellar population (70$\%$), with the remaining dust heating powered by localized embedded sources in NGC\,4565.

The results from this detailed dust energy balance study in NGC\,4565 accords with similar analyses of other edge-on spirals, concluding that a significant fraction of the dust is distributed in such a way that the influence on the overall extinction of a galaxy is negligible but does emit strongly in far-infrared/submm wavebands.  

\section*{Acknowledgements}    
IDL and GG are postdoctoral researchers of the FWO-Vlaanderen.
MB, JF, IDL and JV acknowledge the support of the Flemish Fund for Scientific Research (FWO-Vlaanderen),
in the frame of the research projects no. G.0130.08N and no. G.0787.10N .
The research leading to these results has received funding from the European Community's Seventh Framework Programme (/FP7/2007-2013/) under grant agreement No 229517. 

SPIRE has been developed by a consortium of institutes led by Cardiff University (UK) and including Univ. Lethbridge (Canada); NAOC (China); CEA, OAMP (France); IFSI, Univ. Padua (Italy); IAC (Spain); Stockholm Observatory (Sweden); ISTFC and UKSA (UK); and Caltech/JPL, IPAC, Univ. Colorado (USA). This development has been supported by national funding agencies: CSA (Canada); NAOC (China); CEA, CNES, CNRS (France); ASI (Italy); MCINN (Spain); Stockholm Observatory (Sweden); STFC (UK); and NASA (USA). HIPE is a joint development by the Herschel Science Ground Segment Consortium, consisting of ESA, the NASA Herschel Science Center and the HIFI, PACS and SPIRE consortia.

The analysis in this paper is furthermore based on observations made with the NASA Galaxy Evolution Explorer and the Spitzer Space Telescope. 
GALEX is operated for NASA by the California Institute of Technology under NASA contract NAS5-98034. Spitzer is operated by the Jet Propulsion Laboratory, California Institute of Technology under a contract with NASA.
This publication makes also use of data products from the Sloan Digital Sky Survey (SDSS) and Wide-field Infrared Survey Explorer (WISE). 
Funding for the SDSS and SDSS-II has been provided by the Alfred P. Sloan Foundation, the Participating Institutions, the National Science Foundation, the U.S. Department of Energy, the National Aeronautics and Space Administration, the Japanese Monbukagakusho, the Max Planck Society, and the Higher Education Funding Council for England. The SDSS Web Site is http://www.sdss.org/.
The SDSS is managed by the Astrophysical Research Consortium for the Participating Institutions. The Participating Institutions are the American Museum of Natural History, Astrophysical Institute Potsdam, University of Basel, University of Cambridge, Case Western Reserve University, University of Chicago, Drexel University, Fermilab, the Institute for Advanced Study, the Japan Participation Group, Johns Hopkins University, the Joint Institute for Nuclear Astrophysics, the Kavli Institute for Particle Astrophysics and Cosmology, the Korean Scientist Group, the Chinese Academy of Sciences (LAMOST), Los Alamos National Laboratory, the Max-Planck-Institute for Astronomy (MPIA), the Max-Planck-Institute for Astrophysics (MPA), New Mexico State University, Ohio State University, University of Pittsburgh, University of Portsmouth, Princeton University, the United States Naval Observatory, and the University of Washington.
WISE is a joint project of the University of California, Los Angeles, and the Jet Propulsion Laboratory/California Institute of Technology, funded by the National Aeronautics and Space Administration.

\bsp  

\label{lastpage}

\end{document}